# The Anatomy Spread of Online Opinion Polarization: The Pivotal Role of Super-Spreaders in Social Networks

Yasuko Kawahata [†]

Faculty of Sociology, Department of Media Sociology, Rikkyo University, 3-34-1 Nishi-Ikebukuro,Toshima-ku, Tokyo, 171-8501, JAPAN.
`ykawahata@rikkyo.ac.jp,kawahata.lab3@damp.tottori-u.ac.jp`

**Abstract:** This study delves into the complexities of opinion formation within network environments, focusing on the significant impact of specific entities known as superspreaders. These superspreaders, categorized into three types A, B, and C - play distinct roles in shaping opinions across the network. Superspreader A emerges as a dominant force, characterized by a high z-score indicating its profound influence in dense opinion areas, particularly in local communities and online spaces. This entity can sway opinions even with potentially misleading information, thereby creating filter bubbles and echo chambers. Conversely, Superspreader B possesses a lower z-score and opinion density, operating as a counterbalance to A. B influences opinions in the opposite direction of A, acting as a mitigating factor against the echo chambers and filter bubbles typically fostered by A. Superspreader C, with a high $z-score$ but low opinion density, serves a unique role. It acts as an objective observer, disseminating third-party opinions and functioning akin to media. This entity can either bolster or counteract the influence of A or B, depending on the situation, and is hypothesized to act as a coordinator or fact-checker within the network. The research introduces a confidence coefficient $Dij$ and the z variable to model the behavioral changes of these superspreaders. The study demonstrates how A's influence is pronounced in communities with a high initial trust in its opinions. Finally, we discuss the case of five group dynamics that consider ForgetfulnessFactor and CommunityPersistence, as well as the case of opinion dynamics that include environmental relationships such as future group and location. We also discuss the divergence of opinion motivation in group dynamics for several groups A-E. In particular, we will discuss the conditions that influence group dynamics: environmental variables: are they local, regional, online-dependent, or are they environmentally independent cases? We also conducted initial computational experiments on chronological dependence on, and attachment to, opinions, as well as on forgetting over time. By enhancing the detection of unidirectional communication patterns, this study contributes to network analysis, particularly in safeguarding online communications and understanding social influence dynamics. It provides valuable insights for policymakers, researchers, and social scientists in comprehending information transfer and opinion formation in real-world social networks.

## 1. Introduction

In today's society, the transmission of information and the formation of opinions are extremely important. Social networks, media, and personal interaction influence opinion formation, and these factors can lead to significant changes in politics, economics, and culture. This study explores the dynamics of opinions in complex network environments and the impact of specific factors (superspreaders) on these dynamics.

In this study, we include a $z$ variable at the end to change the behavior of the super-spreader around $z$. Super-spreader A has a high $z$ score under initial conditions and also under conditions of high opinion density in the vicinity. It has a high influence in the community, especially in the local area and within the Internet space. It shall have a high influence on the power of speech and a high efficacy in zones of even higher opinion density, like a risk factor that is prone to constantly generate filter bubbles and echo chambers in its behavior.

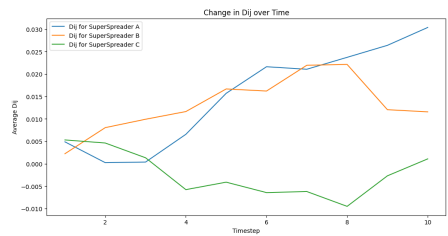

Fig. 1: Case 1:Change in $Dij$ over Time, Super-Spreaders A,B,C

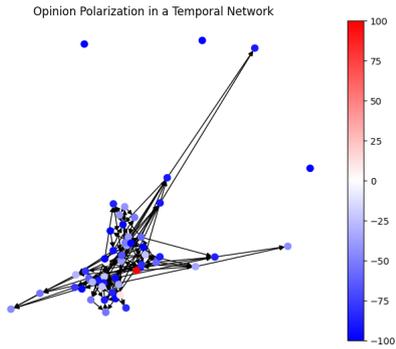

Fig. 2: Case 1:Opinion Polarization in a Temporal Network $n = 50$

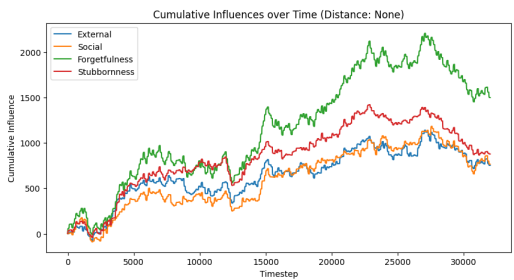

Fig. 3: Case 1:Cumulative Influences over Time, Distance None

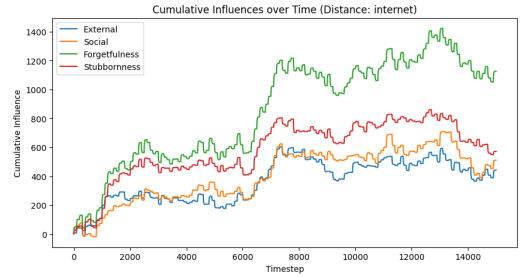

Fig. 4: Case 1:Cumulative Influences over Time, Distance Internet

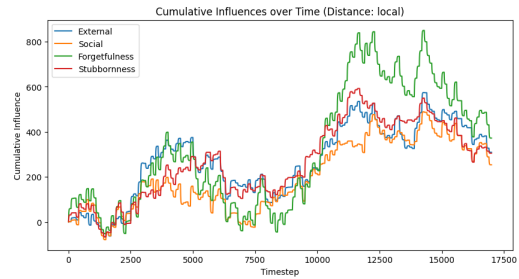

Fig. 5: Case 1:Cumulative Influences over Time, Distance Local

Super-spreader B is initially conditioned to have a Z score about half that of A and also to have opinion density in the vicinity about half that of A. They are particularly influential in their communities, such as in the local area and within the Internet space. The statements are the exact opposite of Super Spreader A, antagonizing Super Spreader A and attempting to change the opinions of clusters influenced by Super Spreader A. It acts on factors that suppress risk factors that tend to constantly generate filter bubbles and echo chambers.

Super Spreader C has a very high Z score under initial conditions but a very low opinion density further out in the neighborhood. The statements are opinions in a completely different direction from those of super-spreaders A and B. They are transmitters of third-party opinions as over-viewers. In other words, it functions like a media. Randomly, it may emphasize the opinion of super-spreader A, the opinion of super-spreader B, or the opinion of another direction. The cluster is assumed to be an objective opinion cluster compared to A and B. It is assumed to be a small cluster but with high opinion launching power.

We would like the following code with additional variables and conditions. Put in a confidence coefficient $D_{ij}$ and a $z$ variable to change the behavior of the super-spreader around $z$. Super-spreader A should have a high $z$ score in the initial condition, plus a high opinion density condition in the vicinity. It is particularly influential in communities, such as in the local area and within the Internet space. In addition, there are many clusters with a high initial value of $D_{ij}$, the coefficient of trust in the opinion of Super Spreader A. These are communities such as fan clubs, religions, and so on. Even if a statement is a lie, the influence of the power of the statement is high and it is easy to take it on board. Furthermore, it shall be highly effective in zones of high opinion density and behaves like a risk factor that is prone to constantly generate filter bubbles and echo chambers.

Super Spreader B is initially conditioned to have a Z score about half that of A, and furthermore, to have opinion density in the vicinity that is also about half that of A. It is particularly influential in communities, such as in the local area and within the Internet space. In addition, many clusters have a high trust coefficient $D_{ij}$ for the opinion of super-spreader B under the initial conditions, while the trust coefficient $D_{ij}$ from the cluster of super-spreader A is low. However, when super-spreader B's statements tend to be emphasized by super-spreader C, the trust coefficient of super-spreader A's cluster itself drops significantly. At this time, the supporters of Super Spreader A flee to Super Spreader B or Super Spreader C.

Super Spreader B speaks the exact opposite of Super Spreader A, antagonizing Super Spreader A and attempting to change the opinions of the clusters affected by Super Spreader A. It acts on factors that suppress risk factors that tend to constantly generate filter bubbles and echo chambers.

Super Spreader C has a very high Z score under initial conditions but a very low opinion density further out in the neighborhood. Super Spreader C has a similar initial confidence factor from Super Spreader B and Super Spreader A. The statements are opinions in a completely different direction from Super Spreader A and B. They are senders of third-party opinions as an overhead person. In other words, it functions like a media. It may randomly emphasize the opinion of super-spreader A, the opinion of super-spreader B, or the opinion of another direction. The cluster is assumed to be a cluster that expresses objective opinions compared to A and B. It is a small cluster but has high opinion launching power. The analysis is based on the hypothesis that the cluster will serve as a coordinator of the opinions of A and B and will function as a fact-checker.

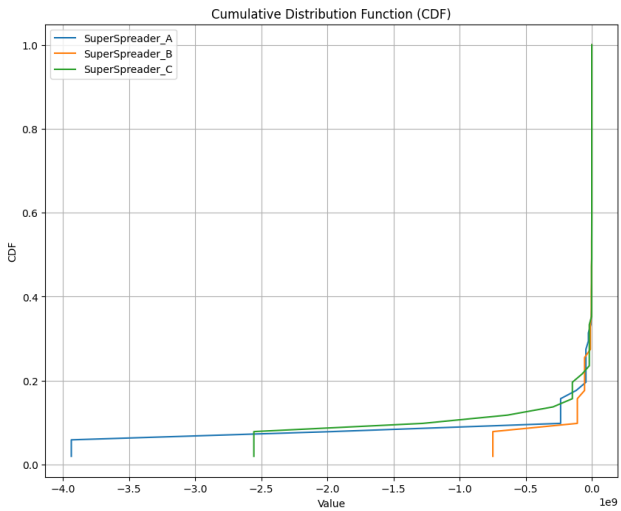

Fig. 6: Case 2:Cumulative Distribution Function (CDF), Super-Spreaders $A,B,C$

As described above, information transmission and opinion formation in a social network depend on the network structure consisting of nodes and edges. Nodes represent individuals or agents, and edges represent relationships between nodes. Due to interactions between nodes, an individual's opinion can change and influence other individuals. Understanding these network dynamics is important for predicting information transfer, assessing social impact, and providing insight into policy decisions.

In this study, we use the following mathematical equations and keywords to simulate and analyze the opinion formation process:

### 1. Network Initialization

We generate an initial network and set the attributes of each node (opinion, robustness, sensitivity to external influences,

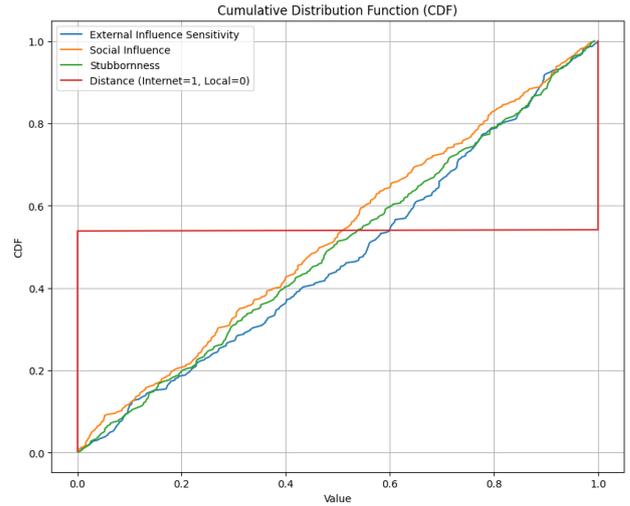

Fig. 7: Case 2:Cumulative Distribution Function (CDF), Type of Opinion

etc.). This defines the starting point for the simulation.
- Node: Represents an individual or agent, each node interacts within the network. - Edges: Represent relationships between nodes and allow for information transfer and propagation of effects. - Attributes: Are characteristics that each node possesses, including opinions, robustness, and sensitivity to external influences.

### 2. Modeling Network Dynamics

At each time step, we model the dynamics of adding and removing edges in the network, as well as changes in individual opinions. This mimics the interaction between nodes and the propagation of opinions.
- Timestep: A discrete unit of time that represents the progression of the simulation. - Adding and Removing Edges: Connections between nodes change over time, forming new relationships or dissolving existing ones. - Individual Opinion Changes: Each node's opinion changes over time and influences other nodes in the network.

### 3. Introduction of a Super-Spreader

At certain time steps (e.g., the 50th step), special super-spreader nodes are introduced. These superspreaders belong to different types and have different influences.
- Superspreader: A specific type of node that serves to facilitate the rapid spread of information. - Type: Indicates the type of superspreader and determines the manner in which they influence information propagation.

# 4. Impact Analysis of Super-Spreaders A, B, and C

After a super-spreader intervenes in the network, their impact is tracked. Understand what role they play in opinion spreading and visualize their effects.

In the vast realm of online communication, identifying unilateral preference (preference) patterns is critical to understanding and mitigating risks such as cyber attacks and information bias. This paper presents a comprehensive approach to analyze and visualize these patterns within social networks. Specifically, we use a directed network model to simulate the flow of information between clusters (groups) with different roles.

The simulation considers the following scenarios:

## 1. Dominant Cluster (A)

This cluster has a strong influence in the network and disseminates information over a wide area.

## 2. Passive Cluster (B)

This is a large but passive cluster that is the primary recipient of information from A. It is the only cluster in the network that is not a dominant cluster.

## 3. Alert Cluster (C)

This cluster monitors A's activities and is alert to its influence. It restricts or blocks the flow of information to protect itself from A's influence.

## 4. Counter Cluster (E)

Takes a stand against A's information and tries to limit its influence.

The key novelty of this study is the combination of dynamic network analysis and speech analysis to reveal the core (core) areas of structural robustness (durability) and influence within the network. 1000 time steps were used to simulate the network and track the evolution of message flow and the emergence of core-periphery structures.

The visualization of the core provides a macroscopic view (the big picture) of the network topology (structure), vividly depicting the resilience of the network and the centrality (importance) of different clusters. Our method captures the expansion of the largest connected components (the largest group of nodes directly or indirectly connected to each other in the network) and identifies important nodes that may act as gatekeepers or influencers in the network.

The results of this study augment the detection of unidirectional communication patterns and provide deep insights into the underlying architecture (underlying structure) that facilitates such interactions. This research contributes to the field of network analysis by demonstrating novel applications in the context of social network simulation that can help protect online communications and provides insight into the role of superspreaders in the spread of opinion. This will allow for a deeper understanding of information communication and opinion formation, and will provide strategic insights into social influence.

This study aims to contribute to our understanding of information transfer and opinion formation in real-world social networks and provide important insights for policy makers, researchers, and the social sciences.

## 1.1 ForgetfulnessFactor and CommunityPersistence Dynamics

Finally, we discuss the case of opinion dynamics, including future groups and environmental relationships such as location, with respect to the case of five group dynamics, taking into account ForgetfulnessFactor and CommunityPersistence. We also discuss the divergence of opinion motivation in the group dynamics of several groups A-E.

# 2. Mainly Method

In today's society, the transmission of information and the formation of opinions are influenced by social networks, media, and personal interaction. This study explores the dynamics of opinions in complex network environments, particularly focusing on the impact of specific factors known as *superspreaders*.

## Modeling Network Dynamics

Let $G = (V, E)$ be a directed graph where $V$ is the set of nodes representing individuals or agents in the network, and $E$ is the set of edges representing relationships or interactions between these nodes. Each node $v \in V$ has an associated opinion $O_v$ which can be influenced by neighboring nodes and external factors.

> Nodes represent individuals or agents, each node interacts within the network.
>
> Edges represent relationships between nodes, allowing for information transfer.
>
> Attributes of nodes include opinions, robustness, and sensitivity to external influences.

## Time-Dependent Network Dynamics

At each time step $t$, the network evolves with the addition or removal of edges and changes in individual opinions.

$$O_v(t+1) = O_v(t) + \Delta O_v(t)$$
$$\Delta O_v(t) = \sum_{u \in \text{Neighbors}(v)} w_{uv} \cdot (O_u(t) - O_v(t))$$

where $w_{uv}$ is the influence weight from node $u$ to node $v$.

## Introduction of Superspreaders

At certain time steps, special nodes called *superspreaders* are introduced into the network. These nodes have a significant impact on the spread of opinions.

## Types of Superspreaders

Superspreader A: High influence in specific clusters with high trust coefficients $D_{ij}$ and opinion density $z$.

Superspreader B: Opposes Superspreader A, attempts to change opinions in influenced clusters.

Superspreader C: Functions as an objective observer, providing third-party opinions.

## Case of Impact Analysis

The impact of superspreaders on the network is analyzed over time, tracking the changes in opinions and the flow of information. This study contributes to understanding the dynamics of information transmission and opinion formation in social networks, highlighting the role of superspreaders in influencing social discourse.

# 3. Network Opinion Polarization Simulation

As a Case Study, we first examine (1) the external influence term of the information, (2) the size of the surrounding community (diffusion of information), (3) the social influence (influencers, or local, general) (4) probability of being influenced by opinions from different clusters (5) computational experiments with distance (score of whether the cluster is Internet-dependent or local).

## 3.1 Model Description

The network is modeled as a directed graph $G = (V, E)$, where $V$ is the set of nodes (agents) and $E$ is the set of directed edges representing connections between nodes.

Each node $i \in V$ has the following attributes:

Opinion $o_i \in [-100, 100]$.

External influence factor $\alpha_i \in [0, 1]$.

Community size (small, medium, large).

Social influence factor $\beta_i \in [0, 1]$.

Susceptibility factor $\gamma_i \in [0, 1]$.

Distance type (internet/local).

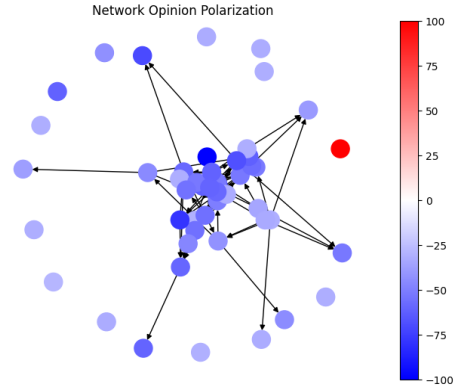

Fig. 8: Network Opinion Polarization $N = 50$

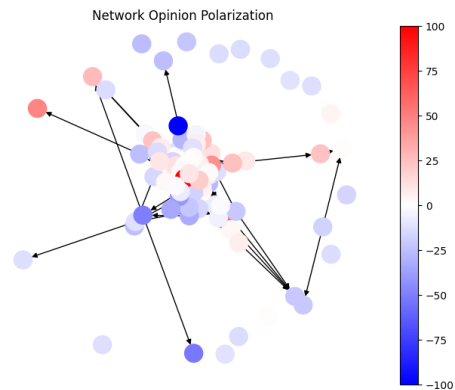

Fig. 9: Network Opinion Polarization $N = 100$

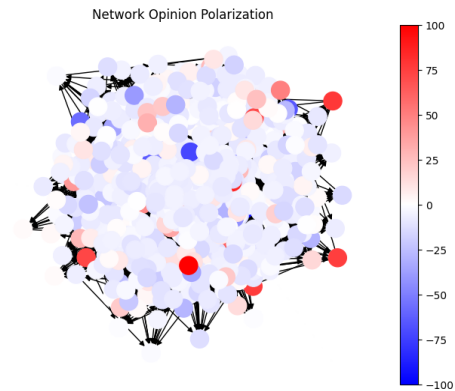

Fig. 10: Network Opinion Polarization $N = 1000$

## 3.2 Network Initialization

The network is initialized with $N = 100$ nodes. Each node is assigned random values for its attributes:

$$o_i \sim \text{Uniform}(-100, 100),$$
$$\alpha_i, \beta_i, \gamma_i \sim \text{Uniform}(0, 1).$$

## 3.3 Link Establishment

Links between nodes are established based on the distance attribute:

For nodes with 'local' distance type, links are formed with probability 0.1.

For nodes with 'internet' distance type, links are formed with probability 0.01.

## 3.4 Opinion Update Function

The opinion of each node is updated considering the social and external influence:

$$o_i^{(t+1)} = (1 - \gamma_i)o_i^{(t)} + \gamma_i \overline{o}_{\text{pred},i} + \alpha_i \epsilon_i, \qquad (1)$$

where $\overline{o}_{\text{pred},i}$ is the average opinion of node $i$'s predecessors and $\epsilon_i \sim \mathcal{N}(0, 1)$ represents random external influence.

## 3.5 Visualization

The simulation is run for a specified number of iterations, and the network's opinion polarization is visualized using a color map ranging from -100 to 100. This model allows the analysis of opinion dynamics and polarization in complex social networks.

To compare and discuss the results for n=50, n=100, and n=1000, we'll consider the social phenomenon, the impact of community size in different environments (local vs. internet), and the consensusbuilding process.

### 1. Social Phenomenon Observation

n=50: The network might represent a small community or group where individual opinions can be heavily influenced by local interactions. The polarization appears moderate, indicating a potential for consensus or balanced discourse. n=100: This could be a mediumsized community. The increased node count may lead to more diverse opinions, but still with a significant potential for local influences to shape consensus. n=1000: A large community, likely representing a broader section of society or a vast online network. The high number of nodes and the dense connections suggest a complex interplay of opinions, possibly leading to echo chambers or fragmented groups with strong polarization.

### 2. Impact of Community Size (Local vs. Internet)

Local Environment: In smaller, local settings (n=50 and n=100), the opinion dynamics are likely to be more homogenous due to frequent and closeknit interactions, and the consensus might be easier to reach. Internet Environment: For larger sizes (n=1000), the internet's broad reach can increase the diversity of opinions and the potential for polarization, making consensus more challenging.

### 3. ConsensusBuilding Consideration

n=50: Smaller groups may reach consensus more readily, but it could be biased by dominant voices due to the higher visibility of individual opinions. n=100: There's a balance between diversity of opinions and the potential for group consensus, which might be ideal for democratic decisionmaking. n=1000: The consensus may require structured mediation and active community management to overcome the inherent noise and potential for misinformation spread due to the vast number of participants and opinions.

The model's parameters (opinion, external influence, community size, social influence, susceptibility, and distance type) play a crucial role in shaping these dynamics. For instance, higher susceptibility and social influence factors might lead to rapid consensus in smaller communities but can cause widespread polarization in larger ones if not managed correctly. Conversely, strong external influences can disrupt local consensus or counteract echo chamber effects in large internet communities, depending on how these influences are perceived and integrated by the individuals within the network.

In conclusion, the visualizations suggest that as community size increases, the complexity of opinion dynamics also rises. This underscores the importance of considering both the microlevel interactions (like individual susceptibility and influence) and macrolevel structures (like community size and connection type) when analyzing social networks and trying to facilitate consensus or understand polarization.

# 4. Local community and Social Ties:Network Opinion Formation Model

The model simulates the evolution of opinions within a network of individuals influenced by their local community and social ties.

## 4.1 Network Initialization

A directed graph $G$ with $N$ nodes is created, where each node represents an individual in the network.

The number of nodes $N$ is set to 50.

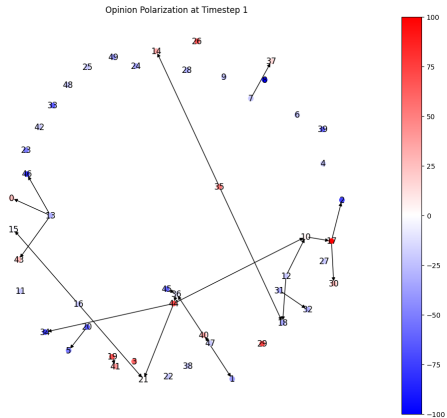

Fig. 11: Network Opinion Polarization $t = 1$

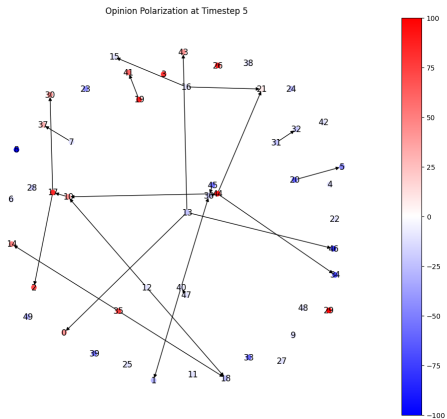

Fig. 12: Network Opinion Polarization $t = 5$

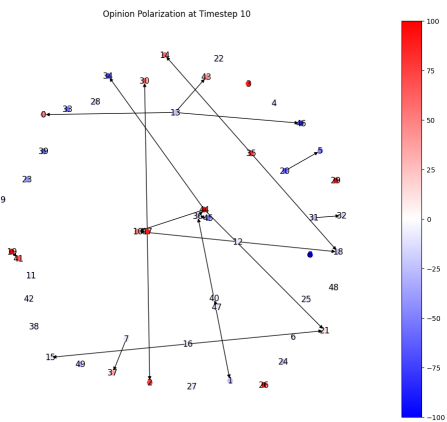

Fig. 13: Network Opinion Polarization $t = 10$

Nodes are initialized with attributes for opinion, community size, social influence, and distance type.

Opinions $o_i$ are randomly assigned from a uniform distribution $U(-100, 100)$.

Community sizes are randomly chosen from {small, medium, large}.

Social influence factors $\beta_i$ are randomly assigned from a uniform distribution $U(0, 1)$.

Distance types are randomly chosen from {internet, local}.

### 4.2 Link Establishment

Edges between nodes are established based on the social influence and community size.

For a node $i$, the influence factor is calculated as the product of the node's social influence $\beta_i$ and a mapping based on community size.

The probability of forming a link is proportional to the influence factor, with local connections being more probable.

### Opinion Formation

At each timestep, the opinion of each node is updated based on the following rules:

For nodes with local distance, the new opinion is the average of the neighbors' opinions weighted by the community influence.

For nodes with internet distance, the new opinion is influenced by a random fluctuation around the current opinion, also weighted by the community influence.

With node colors representing opinions on a blue (negative) to red (positive) scale.

### 1. Social Phenomenon as a Consideration

Timestep 1: The opinions are distributed randomly. This initial condition shows a lack of any significant opinion clusters or consensus. Timestep 5: Some clustering might start to appear, as individuals with similar opinions likely influence each other more strongly, leading to the beginnings of consensus or polarization. Timestep 10: The opinion clusters might become more defined, indicating a stronger polarization within the network. The communities are more likely to be influenced by local majorities or dominant opinions.

## 2. Impact of Community Size (Local vs. Internet) as a Consideration

Local Environment: At earlier timesteps, the local connections could lead to strong clusters of shared opinion due to higher interaction probabilities. This could represent echo chambers where reinforcement of existing opinions occurs. Internet Environment: Over time, the broader reach of the internet connections might introduce more diverse opinions, potentially disrupting local consensus or leading to more widespread but less intense polarization.

## 3. Consensus Formation as a Consideration

Timestep 1: There is likely to be little consensus due to the initial randomness of opinions and lack of established communication patterns. Timestep 5: As nodes influence each other, consensus might begin to form within smaller, more tightlyknit clusters, particularly in local settings. Timestep 10: By this time, some form of consensus could be observed within the network, particularly if certain opinions have become dominant through social influence. However, this consensus might be very polarized, with distinct groups holding opposing views. When interpreting these results, we must consider the underlying mechanisms of the model:Influence Factors: These play a crucial role in the dynamics of opinion formation. A higher social influence factor can lead to quicker consensus but also to stronger polarization.Random Fluctuations: The random fluctuations in opinion for nodes with 'internet' distance can introduce variability, representing realworld factors such as exposure to new information or events.Community Size: The model suggests that individuals in larger communities have more influence, which can lead to a greater impact on the overall opinion landscape. The evolution of the network over time from these images suggests that while local clusters may form quickly, the overall network takes time to reach a broader consensus, which may or may not occur depending on the strength and reach of influential nodes. The model highlights the complexity of opinion dynamics and the significant impact of both local interactions and broader social connections. It also underscores the influence of community size and structure on the opinion formation process within social networks.

## 5. Opinions within Social Network, Taking into Account Individual Predispositions and Interactions

This simulation models the evolution of opinions within a social network, taking into account individual predispositions and interactions.

A directed graph $G$ with $N$ nodes is constructed to represent the social network, where each node corresponds to an individual.

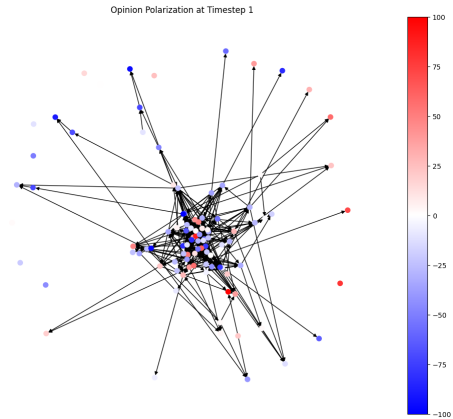

Fig. 14: Network Opinion Polarization $t = 1$

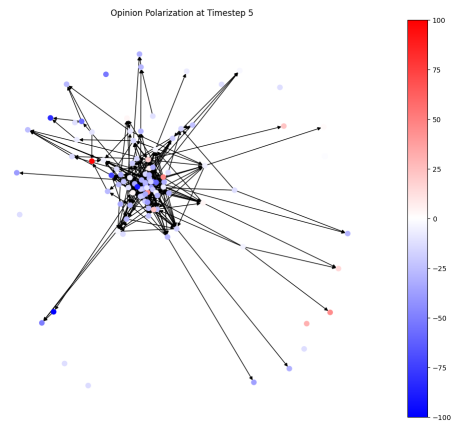

Fig. 15: Network Opinion Polarization $t = 5$

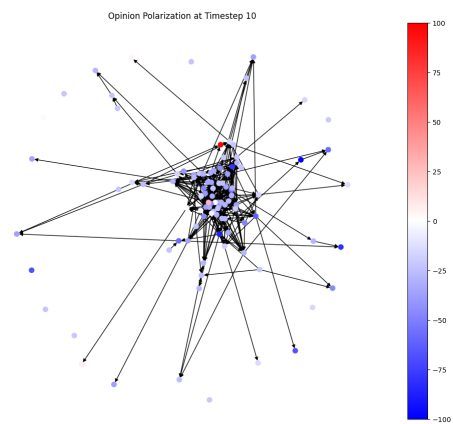

Fig. 16: Network Opinion Polarization $t = 10$

## 1. Update of Node Opinions (Calculation of Opinion Change)

change = (community_opinion−G.nodes[i]['opinion'])·(1−G.nodes[i]['stubbornness'])

- change: Change in node's opinion - community_opinion: Average opinion of nodes within the community - G.nodes[i]['opinion']: Current opinion of node $i$ - G.nodes[i]['stubbornness']: Stubbornness of node $i$ (resistance to change)

## 2. Opinion Change Due to External Influence

external_influence = np.random.randn()·G.nodes[i]['external_influence_sensitivity']

- external_influence: Change in opinion due to external influence - np.random.randn(): Random value from a standard normal distribution - G.nodes[i]['external_influence_sensitivity']: Sensitivity of node $i$ to external influence

## 1. Node-Related Parameters

- $N$: Number of nodes - $G$: Graph data structure (representation of the network) - G.nodes[i]['opinion']: Opinion of node $i$ (initialized with uniform random values) - G.nodes[i]['stubbornness']: Stubbornness of node $i$ (initialized with uniform random values) - G.nodes[i]['external_influence_sensitivity']: Sensitivity of node $i$ to external influence (initialized with uniform random values) - G.nodes[i]['community_size']: Size of the community to which node $i$ belongs (randomly chosen as 'small,' 'medium,' or 'large') - G.nodes[i]['social_influence']: Social influence of node $i$ (initialized with uniform random values) - G.nodes[i]['distance']: Distance attribute of node $i$ (randomly chosen as 'internet' or 'local')

## 2. Parameters Related to Graph Linkage

- Linkages between nodes are set with different probabilities based on the distance attribute. - Probability of connection between nodes within the local community: 0.1 - Probability of connection between nodes within the internet community: 0.01

## 3. Parameters Related to Simulation

- timesteps: Number of simulation timesteps (set to 10 in this context)

  The number of nodes $N$ is set to 100.

  Each node is initialized with the following attributes:

  – Opinion $o_i$ drawn from a uniform distribution $U(-100, 100)$.

  – Stubbornness factor $s_i$ drawn from $U(0, 1)$.

  – Sensitivity to external influence $e_i$ drawn from $U(0, 1)$.

  – Community size $c_i$ chosen randomly from {small, medium, large}.

  – Social influence factor $\beta_i$ drawn from $U(0, 1)$.

  – Distance type $d_i$ chosen randomly from {internet, local}.

### Link Formation

Connections between individuals are established based on their distance types:

  For local nodes, links are formed with a probability of 0.1.

  For nodes connected through the internet, the probability is 0.01.

### Opinion Update

At each timestep, individuals update their opinions based on social influences and personal attributes:

  The new opinion $o'_i$ is calculated by considering the average opinion of neighboring nodes, the individual's current opinion, and their stubbornness.

  If the node is influenced by the internet, a random external influence term is added, weighted by the individual's sensitivity to external influence.

  The simulation runs for a specified number of timesteps:

  At each timestep, opinions are updated using the `update_opinions` function.

  The network is visualized after each update, showing opinion polarization.

## 1. Social Phenomenon

Timestep 1: The network shows a random distribution of opinions, indicating the initial state before interactions have significantly influenced the opinions. Timestep 5: Some clusters may begin to emerge as individuals influence one another, possibly leading to the formation of opinion groups. Timestep 10: The network may display more pronounced clusters or even further polarization, with groups of nodes sharing similar opinions, separated from others with opposing views.

## 2. Community Size and Environment Impact (Local vs. Internet)

Local Environment: Initially, local connections could lead to quicker and stronger opinion convergence within those subgroups, potentially creating echo chambers. Internet Environment: Over time, the larger, more diverse influence of internet connections may introduce a variety of opinions, potentially disrupting local consensus or leading to more moderate, widespread opinions.

### 3. Consensus Formation

Timestep 1: Consensus is unlikely as the network is just beginning to interact, and opinions are randomly distributed. Timestep 5: Some local consensus may be emerging within tightly connected clusters, but a global consensus is still forming. Timestep 10: By this point, the network may have reached a form of dynamic equilibrium, where some form of consensus or stable disagreement may have formed, reflecting realworld scenarios of social opinion dynamics.

In analyzing these images, we can infer the following:

### Stubbornness Factor

Nodes with a higher stubbornness factor will be less likely to change their opinions, potentially acting as anchors within their respective clusters.

### External Influence Sensitivity

Nodes with a high sensitivity to external influence will have more fluctuating opinions, introducing variability and potentially preventing the formation of a strong consensus.

### Social Influence

Nodes with a higher social influence factor will have a more significant impact on their neighbors, possibly leading to the formation of opinion leaders or influencers within the network.

The simulation provides insights into how individual attributes and network structure can lead to complex patterns of opinion formation and polarization over time. It demonstrates the importance of considering multiple factors in understanding social dynamics, including personal traits, local community bonds, and broader social connectivity.

# 6. Each node within the clusters:Opinion Dynamics in Network Simulation

Here, a new attribute called opinion is added to the node to model how opinions change at each time step. It also includes the stubbornness of opinion, $external_influence_sensitivity$, $community_size$, $social_influence$, and distance (Internet dependency or local) distance.

## 1. Update of Node Opinions (Calculation of Opinion Change)

change = (community_op−G.nodes[i]['op'])·(1−G.nodes[i]['stubb'])

- change : Change in node's opinion - community_opinion : Average opinion of nodes within the community - G.nodes[i]['opinion']: Current opinion of node $i$ - G.nodes[i]['stubbornness']: Stubbornness of node $i$ (resistance to change)

## 2. Opinion Change Due to External Influence

external_influence = np.random.randn() · G.nodes[i]

- external_influence: Change in opinion due to external influence
- np.random.randn(): Random value from a standard normal distribution - G.nodes[i]['external_influence_sensitivity']: Sensitivity of node $i$ to external influence

## 1. Node-Related Parameters

- $N$: Number of nodes - $G$: Graph data structure (representation of the network) - G.nodes[i]['opinion']: Opinion of node $i$ (initialized with uniform random values) - G.nodes[i]['stubbornness']: Stubbornness of node $i$ (initialized with uniform random values) - G.nodes[i]['external_influence_sensitivity']: Sensitivity of node $i$ to external influence (initialized with uniform random values) - G.nodes[i]['community_size']: Size of the community to which node $i$ belongs (randomly chosen as 'small,' 'medium,' or 'large') - G.nodes[i]['social_influence']: Social influence of node $i$ (initialized with uniform random values) - G.nodes[i]['distance']: Distance attribute of node $i$ (randomly chosen as 'internet' or 'local')

## 2. Parameters Related to Graph Linkage

- Linkages between nodes are set with different probabilities based on the distance attribute. - Probability of connection between nodes within the local community: 0.1 - Probability of connection between nodes within the internet community: 0.01

## 3. Parameters Related to Simulation

- timesteps: Number of simulation timesteps (set to 10 in this context)

The network is initialized with clusters, each with a predefined size and number of messages.

Cluster sizes: $A = 100, B = 100, C = 50, D = 60, E = 10$.

Messages per cluster: $A = 500, B = 100, C = 100, D = 50, E = 100$.

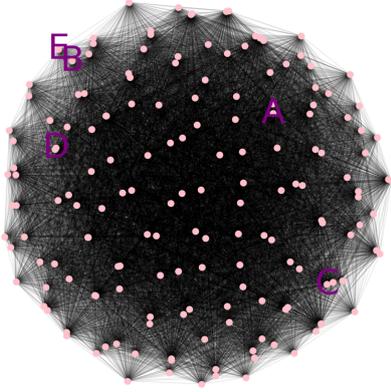

Fig. 17: Largest Connected Component with k-core (k=2, 6, 23) at Timestep $t = 10$

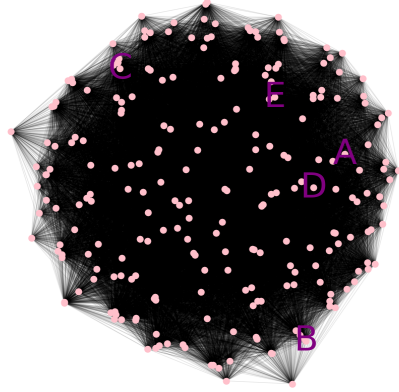

Fig. 19: Largest Connected Component with k-core (k=2, 6, 23) at Timestep $t = 100$

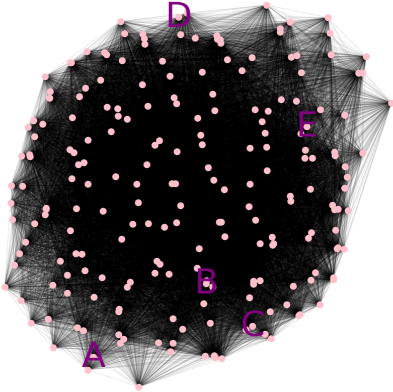

Fig. 18: Largest Connected Component with k-core (k=2, 6, 23) at Timestep $t = 50$

Each node within the clusters is initialized with the following attributes:

$$\text{Opinion} \sim U(-100, 100),$$
$$\text{Stubbornness} \sim U(0, 1),$$
$$\text{External Influence Sensitivity} \sim U(0, 1),$$
$$\text{Community Size} \in \{\text{small, medium, large}\},$$
$$\text{Social Influence} \sim U(0, 1),$$
$$\text{Distance} \in \{\text{internet, local}\}.$$

## Opinion Update

At each timestep, the network's nodes update their opinions based on the following factors:

Forgetfulness factor $\phi = 0.1$, applied selectively based on the node's distance.

Community persistence $\psi = 0.5$, enhancing opinion retention for nodes with high social influence.

Negative opinions are pruned over time within local communities.

Positive opinions lead to active connection establishment in the network.

The largest connected component and its k-core are visualized, highlighting different levels of connectivity and the presence of clusters.

The $k$-core is computed for $k = 2, 6, 23$.

Node positions are determined using a force-directed layout.

Clusters are labeled based on the dispersion of their node positions.

The simulation runs over 100 timesteps, updating opinions and visualizing the network at intervals.

### 1. Social Phenomenon Observation

Over time, the network seems to coalesce into distinct clusters. At early timesteps (10), the network is less structured, but by timestep 50 and more so by 100, the clusters become more pronounced, indicating that over time, individuals within the network tend to form tighter communities of shared opinions or interests.

### 2. Influence of Media

The impact of media, possibly represented by external influences in the simulation, can be seen in how nodes that are more sensitive to external influences (i.e., those with an

'internet' distance attribute) might shift their opinions more significantly over time. This could lead to opinion changes or the introduction of new topics within clusters.

### 3. Consensus Formation

Consensus within the network might form within individual clusters where nodes share similar opinions, but across the entire network, consensus seems unlikely due to the presence of multiple distinct clusters. This suggests that while localized agreement may be possible, a global consensus is challenging due to the diversity of opinions and interests.

### 4. ClusterSpecific Observations (A, B, C, D, E)

Each cluster's behavior can be influenced by its initial size and message count. Smaller clusters (like E) may show tighter cohesion due to their size, while larger clusters (A and B) have more potential for internal variability. The evolution of these clusters over time may reflect how different communities adapt and interact within the larger social network.

### 5. Community Size and Environment Impact (Local vs. Internet)

Local nodes may display stronger connections within their immediate community, reflecting localized consensus or echo chambers. Internet nodes, on the other hand, could have more dispersed opinions due to their broader exposure and potentially less community cohesion. Over time, the local nodes may maintain their opinions more robustly (due to higher community persistence), while internet nodes may exhibit more opinion volatility.

In conclusion, the network simulation reveals the complex dynamics of social interactions and opinion formation. The model shows that individual predispositions (such as stubbornness and sensitivity to external influences), along with communication patterns (local vs. internet), significantly affect how opinions evolve and how communities within the network develop and interact with each other. The visualization at different timesteps highlights the importance of these factors in shaping the social structure of the network.

### Cluster A

Initially, this cluster starts with the largest size and the highest number of messages. This suggests a highly active and influential community within the network. Over time, we expect Cluster A to form a dense core in the network due to its high activity level, potentially attracting more nodes into its influence sphere. By timestep 100, Cluster A may have consolidated its opinions, becoming a dominant voice in the network, possibly due to its initial size and message volume.

### Cluster B

Similar in initial size to Cluster A, Cluster B has fewer messages, indicating a potentially less active or less influential group initially. The evolution of Cluster B might show a slower consolidation of opinions compared to Cluster A, with a less dense core due to the reduced number of messages. By timestep 100, Cluster B is likely to form its own distinct community within the network, but it may not be as central or influential as Cluster A.

### Cluster C

Being smaller in size, Cluster C is likely to be more nimble in its opinion dynamics, with individual nodes possibly playing a more significant role due to the smaller community size. The moderate number of messages suggests a balanced level of activity which may lead to moderate influence within the network. By timestep 100, Cluster C might exhibit a more tightlyknit community with more uniform opinions but might be somewhat isolated from larger clusters due to its smaller size.

### Cluster D

With a size larger than Cluster C but smaller than A and B, and with the fewest messages, Cluster D may exhibit a mix of dynamics between large and small clusters. Cluster D's influence on the network may grow more through the accumulation of shared opinions over time rather than through initial message volume. By timestep 100, Cluster D is likely to have formed a distinct subgroup within the network, possibly with internal subclusters due to the diversity created by fewer messages.

### Cluster E

As the smallest cluster, Cluster E's behavior is likely to be highly cohesive due to the ease of reaching consensus in small groups. Despite having a high message count relative to its size, Cluster E may struggle to exert influence on the larger network due to its limited size. By timestep 100, Cluster E may remain a highly cohesive but peripheral group within the network, potentially acting as a niche community with specialized opinions or interests.

These speculations consider the nodes' opinion dynamics, stubbornness, external influence sensitivity, community size, social influence, and distance type as provided in the simulation model. Over time, larger clusters with more messages may tend to become influential opinion leaders, while smaller clusters might develop strong internal consensus but have limited impact on the network as a whole. The interplay between local and internet nodes within each cluster also affects the cohesion and influence of each cluster.

To delve into a detailed consideration of the clusters with respect to the attributes Stubbornness, External Influence Sensitivity, Community Size, Social Influence, and Distance (internet, local), we must consider the roles these attributes play within the simulation's model and how they might manifest differently in each cluster based on the visualizations you've provided.

### Stubbornness

Stubbornness reflects an individual node's resistance to changing its opinion. In a cluster with high average stubbornness, we'd expect to see less fluctuation in opinions over time. Such clusters might maintain their initial position more robustly, potentially acting as opinion anchors within the network.

### External Influence Sensitivity

This attribute determines how susceptible a node is to external information, which could represent media influence or other external factors. Clusters with high sensitivity might show more dynamic changes in opinion, reflecting the evolving nature of the information they are exposed to. Nodes with internet connections are expected to have higher sensitivity, which could lead to more varied and rapidly changing opinions within those clusters.

### Community Size

The community size attribute likely correlates with the number of nodes within a cluster and the density of connections. Larger communities might have more diverse opinions and a greater potential for internal subclustering. Smaller communities might exhibit tighter cohesion and more uniform opinion dynamics due to the stronger influence nodes exert on each other.

### Social Influence

High social influence suggests that certain nodes are more impactful in swaying the opinions of their neighbors. Clusters with nodes of high social influence might centralize around key influential nodes, potentially leading to more pronounced opinion leadership. This could lead to the formation of more distinct opinion clusters if these influencers promote consensus or further polarization if they have opposing views.

### Distance (internet, local)

The distance attribute influences the likelihood of forming connections. Nodes with 'local' distance might be more insulated, leading to clusters that are more isolated and potentially more homogeneous in opinion. On the other hand, 'internet' nodes, with their broader connectivity, might result in clusters that are more exposed to external influences and have a more distributed opinion landscape.

## Considering the visualizations at timesteps 10, 50, and 100

Cluster A Given its size, Cluster A might display a wide range of opinions. The evolution from timestep 10 to 100 could show how dominant opinions emerge or how the cluster reacts to external influences if internet connections are prevalent. Cluster B Similar in size to Cluster A, Cluster B's dynamics would be interesting to compare with A. If B has a higher average stubbornness or lower external influence sensitivity, it might show less opinion change over time. Cluster C Being smaller, Cluster C's dynamics are more influenced by individual nodes. A few nodes with high social influence could significantly shape the cluster's opinion trajectory. Cluster D This cluster's behavior might be a mix between A/B and C, with enough size to have diversity but not so large as to prevent a few nodes from having a substantial impact. Cluster E The smallest cluster, E, might show the most cohesive opinion dynamics. If it has a high average stubbornness and social influence, along with local distance, it could act as a very stable and insulated community within the network.

Each cluster's evolution will also depend on the initial conditions set by the simulation's random attribute assignments. The interactions between these attributes and the network's topology will ultimately shape the clusters' trajectories over the course of the simulation.

To delve into a detailed consideration of the clusters with respect to the attributes Stubbornness, External Influence Sensitivity, Community Size, Social Influence, and Distance (internet, local), we must consider the roles these attributes play within the simulation's model and how they might manifest differently in each cluster based on the visualizations you've provided.

### The cumulative distribution function (CDF)

Results shows the distribution of several attributes within the network Stubbornness, External Influence Sensitivity, Social Influence, and a binary representation of Distance (where Internet is 1, Local is 0). To provide a detailed consideration for each cluster regarding these attributes, we would ideally need more specific data or visualizations that separate the clusters. However, based on the CDF and understanding of the model, here is a general analysis

### Stubbornness

This attribute's CDF indicates a fairly even distribution across the nodes, with a slightly steeper curve in the midrange values. This suggests that most nodes are moderately stubborn, with fewer nodes at the extremes (either very flexible or very stubborn). In a cluster context, a cluster with a higher median stubbornness would be less likely to change its overall opinion over time, potentially resisting external influences.

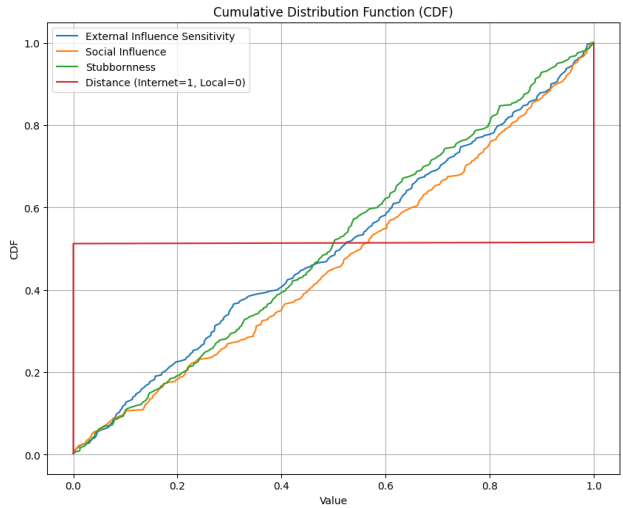

Fig. 20: Cumulative Distribution Function (CDF)

### External Influence Sensitivity

The External Influence Sensitivity has a similar distribution to Stubbornness, but it appears to have a slightly higher concentration of nodes with mid to high sensitivity values. Clusters that have a higher average sensitivity may exhibit greater opinion fluctuation in response to external information, particularly if they have more internetconnected nodes.

### Social Influence

The Social Influence has a noticeably steeper curve at the lower values, indicating that a larger proportion of the network has lower social influence. This suggests that within clusters, influential nodes (those with high social influence) are fewer but may significantly impact the opinion dynamics by potentially swaying the opinions of less influential nodes.

### Distance (Internet=1, Local=0)

The Distance attribute shows a binary distribution, as it's a categorical variable. If the CDF for Distance sharply increases at the value of 1, this would indicate that there are many more internet nodes than local ones. Clusters predominated by internet nodes would be more open to external influence and may have more diffuse boundaries due to the higher connectivity.

To analyze each cluster (A, B, C, D, E) specifically, you would need to look at the distribution of these attributes within each cluster. For example, A cluster with higher stubbornness may be more resistant to change, maintaining its core opinions over time. A cluster with high external influence sensitivity might reflect the broader social trends and changes, showing a greater diversity of opinions. Clusters with a higher proportion of socially influential nodes might show a more hierarchical structure, where certain nodes have a larger impact on the cluster's opinion dynamics. The distance attribute affects how clusters interact with each other and the rest of the network. Clusters with more local nodes might form tightknit communities with more intense but localized interactions, whereas clusters with more internet nodes might be more involved in broader networkwide opinion exchanges. For a comprehensive analysis, the CDFs for each attribute should be compared cluster by cluster, examining how the distributions differ and potentially correlate with the cluster's behavior observed in the simulation. This would provide insights into how these attributes influence the formation and evolution of opinions within and between the clusters over time.

## 7. Each node within the clusters: Opinion Dynamics in Network Simulation

Here, a new attribute called opinion is added to the node to model how opinions change at each time step. It also includes the stubbornness of opinion, $external_{influence_s}ensitivity$, $community_size$, $social_{influence}$, and distance (Internet dependency or local) distance.

### 1. Update of Node Opinions (Calculation of Opinion Change)

$new_opinion$ =
local: (opinion(neighbors) /
count(neighbors)) ×
$community_{i}nfluence$
internet: $current_opinion + random_variation$ ×
$community_{i}nfluence$ - new_opinion: New opinion of node $i$ - $\sum_{\text{neighbor} \in \text{predecessors}(i)}$ G.nodes[neighbor]['opinion']: Sum of opinions of predecessor nodes of node $i$ - count of predecessors($i$): Number of predecessor nodes of node $i$ - community_influence: Influence from the community to which node $i$ belongs - G.nodes[$i$]['distance']: Distance attribute of node $i$ ('local' or 'internet')

### 2. Example of Opinion Update (As part of the simulation)

$$\text{new\_opinions}[i] = \text{new\_opinion} \cdot \sin\left(\frac{\text{timestep}}{10}\right)$$

- new_opinions[$i$]: New opinion of node $i$ (varies with simulation timestep) - $\sin\left(\frac{\text{timestep}}{10}\right)$: Variation using the sine function based on the timestep

## 1. Node-Related Parameters

- $N$: Number of nodes (50) - $G$: Graph data structure (DiGraph) - G.nodes[i]['opinion']: Opinion of node $i$ (initialized with uniform random values) - G.nodes[i]['community_size']: Size of the community to which node $i$ belongs (randomly chosen as 'small,' 'medium,' or 'large') - G.nodes[i]['social_influence']: Social influence of node $i$ (initialized with uniform random values) - G.nodes[i]['distance']: Distance attribute of node $i$ (randomly chosen as 'internet' or 'local')

## 2. Parameters Related to Graph Linkage

- Node linkages are probabilistically set based on community membership and social influence. - influence_factor = community_influence_mapping[G.nodes[i]['community_size']] · G.nodes[i]['social_influence']: Influence factor for link establishment - probability = $\frac{\text{influence\_factor}}{100}$: Probability variable for probabilistic link setting - G.nodes[i]['distance']: Adjustment of probability based on node $i$'s distance attribute - np.random.rand(): Generates a random value in the range $[0, 1]$

## 3. Parameters Related to Simulation

- timesteps: Number of simulation timesteps (10)

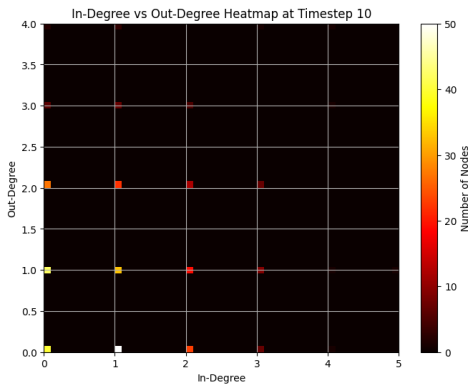

Fig. 21: In-Degree vs Out-Degree Heatmap at Timestep $t = 10$

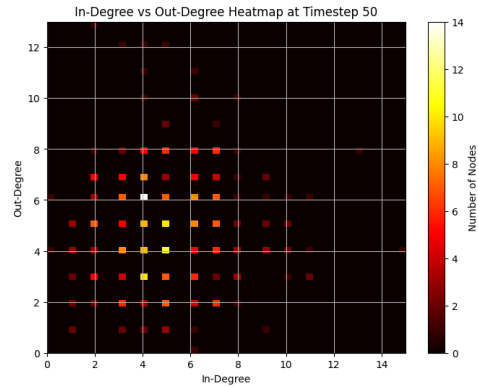

Fig. 22: In-Degree vs Out-Degree Heatmap at Timestep $t = 50$

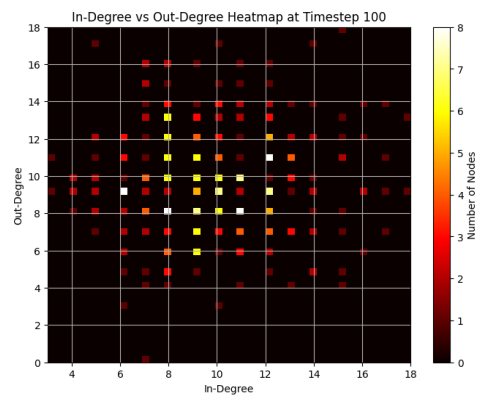

Fig. 23: In-Degree vs Out-Degree Heatmap at Timestep $t = 100$

## 1. Social Phenomenon

Early Stage (Timestep 10) The network is in its formative stage. Relationships are emerging, but the structure is not yet fully established. Opinions are likely to be more diverse. MidStage (Timestep 50) Relationships and opinion clusters become more defined. Social influence and stubbornness contribute to the formation of opinion groups. Late Stage (Timestep 100) Opinion clusters are likely solidified, and external influence sensitivity may cause clusters to adapt or resist external information depending on their exposure (internet vs. local).

## 2. Media Influence

Early Stage Nodes with high external influence sensitivity may significantly alter their opinions due to external (media) information. MidStage Media's impact might solidify or disrupt emerging clusters, depending on whether the information reinforces or contradicts prevailing opinions. Late Stage Media may continue to influence internetconnected nodes, but stubbornness and community persistence could mitigate this influence, stabilizing the clusters.

## 3. Consensus Formation

Early Stage Consensus is unlikely as the network is still forming. Social influence plays a crucial role in shaping early opinion trends. MidStage Subconsensus within clusters may be achieved, especially in communities with high stubbornness and social influence. Late Stage Global consensus across the network is challenging, but stable local consensuses within clusters are likely, shaped by community size and the stubbornness of nodes.

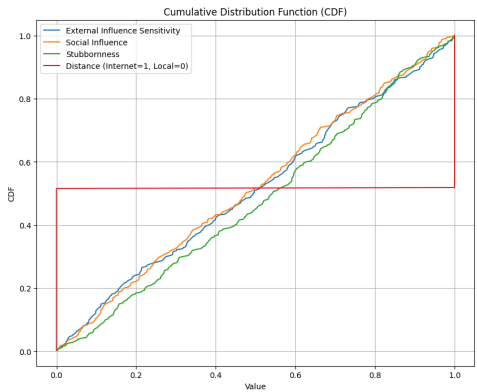

Fig. 24: Cumulative Distribution Function (CDF)

## 4. AttributeSpecific Observations

Stubbornness More stubborn clusters resist opinion change over time, retaining initial biases and possibly leading to polarization. External Influence Sensitivity Clusters with higher sensitivity are more dynamic in their opinion change, reflecting a broader range of external influences. Community Size Larger clusters exhibit a wider range of opinions, whereas smaller clusters show more uniformity and possibly stronger internal consensus. Social Influence Nodes with higher social influence act as opinion leaders, significantly impacting their clusters' opinion dynamics. Distance (Internet vs. Local) Internet nodes are more exposed to external influences, potentially leading to opinion volatility, while local nodes exhibit more opinion stability due to less exposure.

## 5. Influence of Community Size and Environment

Local Environment Clusters with predominantly local nodes may develop strong, stable internal opinions, with high community persistence reinforcing existing opinions. Internet Environment Clusters with a high proportion of internet nodes are more susceptible to external influences, which could lead to more significant opinion shifts and a less stable cluster consensus.

The heatmap visualizations at each timestep indicate the degree distribution within the network, with the concentration of points away from the origin (0,0) suggesting the formation of a hierarchical structure or influential nodes. The CDF plots reveal the distribution of node attributes, which influence how nodes interact and form opinions. As the simulation progresses, we would expect the degree distribution to become more dispersed if new connections form or more concentrated if the network stratifies around influential nodes. The attributes of stubbornness, external influence sensitivity, social influence, and distance shape these dynamics, influencing the network's evolution over time.

## 8. Description of the Opinion Spread Simulation Model

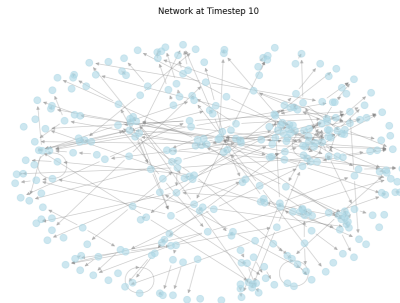

Fig. 25: Cumulative Influences over Time $t = 10$

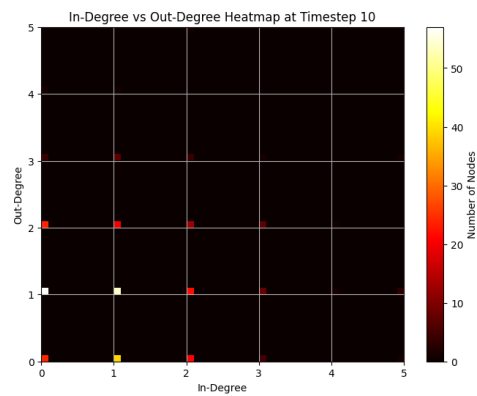

Fig. 26: In-Degree vs Out-Degree Heatmap at Timestep $t = 10$

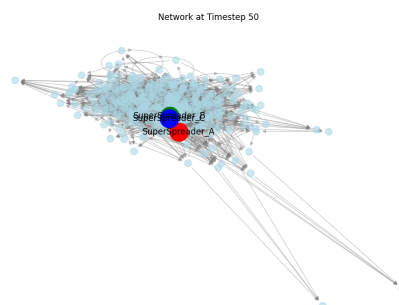

Fig. 27: Cumulative Influences over Time $t = 50$

### 8.1 Model Overview

This model simulates the spread of opinions in a network, with a special focus on the role of *superspreaders* in influenc-ing the opinions of other nodes in the network. The simulation is structured around a directed graph, where each node represents an individual with certain attributes and edges represent the influence relationships.

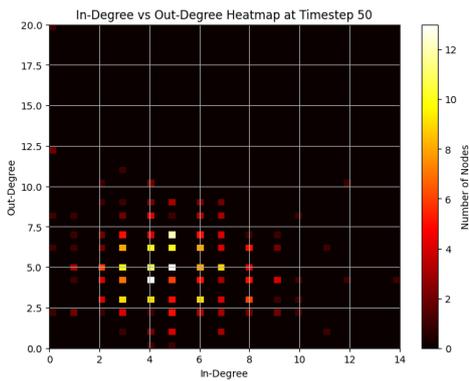

Fig. 28: In-Degree vs Out-Degree Heatmap at Timestep $t = 50$

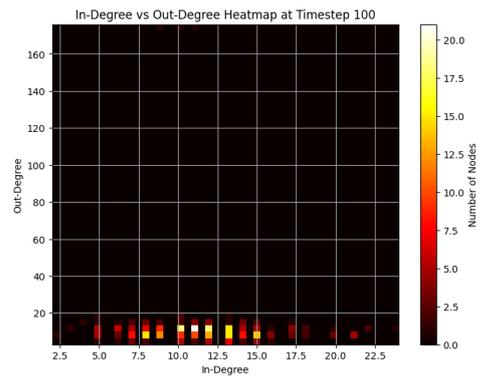

Fig. 30: In-Degree vs Out-Degree Heatmap at Timestep $t = 100$

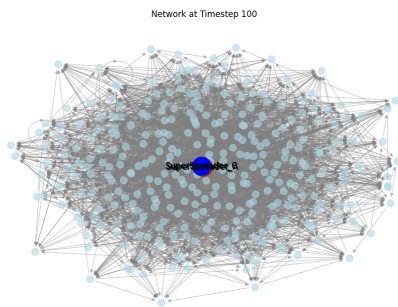

Fig. 29: Cumulative Influences over Time $t = 100$

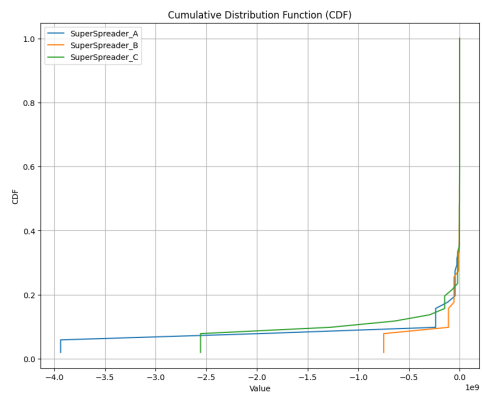

Fig. 31: Cumulative Distribution Function (CDF)

### 8.2 Superspreaders Initialization

Superspreaders are special nodes in the network with significant influence. There are three types of superspreaders:

**SuperSpreader_A**: Represents local media with a randomly chosen opinion from {-100, 0, 100}, and is marked with the attribute *distance* as 'local'.

**SuperSpreader_B**: Represents internet media with a similar opinion range and marked as 'internet'.

**SuperSpreader_C**: Represents a contrarian opinion leader, also with a similar opinion range but focused on opposing the majority opinion.

### 8.3 Opinion Spread Mechanism

The spread of opinions depends on the type of superspreader:

**Local Media**: Influences nearby nodes (local).

**Internet Media**: Has a random influence across the network.

**Contrarian Leader**: Adopts an opinion opposite to the majority and influences accordingly.

Opinions are updated based on a probability and a simple averaging mechanism.

### 8.4 Network Update and Simulation

The network evolves over time with changes in connections and node opinions. At a specific timestep (e.g., 50), superspreaders are introduced. The network is updated at each timestep, considering factors like forgetfulness and social influence.

The model includes functions for visualizing the network, the degree heatmap, and the cumulative distribution function (CDF) of various attributes.

### 8.5 Parameters and Functions

**sizes**: Dictionary specifying the size of different clusters in the network.

**messages**: Dictionary defining the number of messages or influences for each cluster.

**update_network**: Function to update the network at each timestep, influencing node opinions and connections.

**plot_network**: Visualizes the network at a given timestep, highlighting superspreaders.

**plot_degree_heatmap**: Generates a heatmap of the in-degree vs out-degree of nodes.

**plot_cdf**: Plots the cumulative distribution function for given data.

## 8.6 Timestep 10

### 1. Social Phenomenon Observation

The network at Timestep 10 shows a relatively dispersed set of nodes with several connections, suggesting that some communication or influence is happening, but there is no clear pattern of a dominant cluster or central hub. This could represent an early stage in the spread of information or opinions within a social network, where influence has yet to be centralized.

### 2. Consensus Formation

At this early stage, there is unlikely to be any significant consensus. The network lacks strong centrality, which would be necessary to drive a consensus. The process of opinion formation and convergence is probably just beginning, with several nodes acting independently.

### 3. Cluster (A, B, C) Observation

With no clear distinction between clusters A, B, and C at this stage, it's difficult to comment on their specific dynamics. However, given the code provided, we can speculate that these clusters may have varied in terms of stubbornness, external influence sensitivity, and social influence, which would affect how they evolve over time.

### 4. Impact of Community Size Environment (Local, Internet)

At Timestep 10, the effect of community size (i.e., local versus internet) is not yet apparent in the network structure. The local nodes might be expected to form more tightlyknit communities due to physical proximity, while internet nodes could be more widely dispersed. However, given the early stage of the network's evolution, these patterns may not have emerged yet.

Now, examining the InDegree vs OutDegree Heatmap at Timestep 10 The heatmap shows a concentration of nodes with low indegrees and outdegrees, which is consistent with the early stage of network formation where few nodes have become significant broadcasters or recipients of information.

And looking at the Cumulative Distribution Function (CDF) for Superspreaders A, B, and C The CDFs for the superspreaders are not provided for Timestep 10 in the images. Superspreaders are introduced at Timestep 50, so their influence is not present at Timestep 10.

To move forward with a more comprehensive analysis, we would need to display and analyze the network states at Timesteps 50 and 100, as well as any relevant CDF plots and indegree vs outdegree heatmaps for those timesteps. If you could provide the images for Timesteps 50 and 100, we can continue with the analysis.

## 8.7 Timestep 50

### 1. Social Phenomenon Observation

By Timestep 50, the network has evolved with more evident central nodes, labeled as '$SuperSpreader_A$' and '$SuperSpreader_B$'. Their presence indicates that certain nodes (superspreaders) have become significant influencers within the network, potentially representing media outlets or influential individuals in a social context. The network topology has shifted from a dispersed structure to a more centralized one around these superspreaders.

### 2. Consensus Formation

The emergence of superspreaders suggests that the network may be moving towards areas of localized consensus, particularly around '$SuperSpreader_A$' and '$SuperSpreader_B$'. These nodes seem to be broadcasting to a wide audience, possibly swaying the overall network's opinions and leading to the formation of larger clusters of shared beliefs or opinions.

### 3. Cluster (A, B, C) Observation

Clusters A and B are now visible and centered around the superspreaders, reflecting their influence. Cluster C's status isn't clear from the image, but it might be expected to have a contrarian stance given the context provided. The clusters' dynamics are likely to be heavily influenced by the nature of the superspreaders' opinions and their respective reach within the network.

### 4. Impact of Community Size Environment (Local, Internet)

The superspreaders' influence seems to transcend local and internet boundaries, indicating that messages are reaching both local and internetconnected nodes. However, the extent of influence may vary, with local nodes being more intensely affected by '$SuperSpreader_A$' (designated as local media in your code) and a wider but less dense influence from '$SuperSpreader_B$' (an internet media node).

Looking at the InDegree vs OutDegree Heatmap at Timestep 50 There is a more diverse distribution of indegrees and outdegrees, with some nodes (likely the superspreaders) having a high outdegree, indicating their role in spreading information or opinions. And considering the Cumulative Distribution Function (CDF) plot The CDF plot for the superspreaders is not specifically tied to Timestep 50, but it shows a steep curve at the higher opinion values for all three superspreaders, indicating that their opinions may be extreme and have a significant impact on the network's dynamics.

This analysis provides a snapshot of the network's evolution at a critical juncture, after the introduction of superspreaders but before their full influence has been realized. To continue with a comprehensive review, we would next examine the network state at Timestep 100 to see the final or nearfinal state of the network after the superspreaders have had more time to exert their influence.

## 8.8 Timestep 100

### 1. Social Phenomenon Observation

The network has become highly centralized around the superspreaders, with '$SuperSpreader_A$' appearing to have the most prominent position. This reflects a social phenomenon where certain individuals or entities have become dominant opinion leaders, shaping the network's communication flow.

### 2. Consensus Formation

The centralization around the superspreaders suggests that their influence could lead to a stronger consensus within the network, as they disseminate their opinions more widely. The fact that there are multiple superspreaders, however, might mean that there are several competing consensus clusters, especially if the superspreaders represent different opinions or interests.

### 3. Cluster (A, B, C) Observation

The structure of the network suggests that clusters A, B, and C might now be heavily influenced by the superspreaders. The cluster around '$SuperSpreader_A$' is particularly pronounced, indicating that this entity's influence is widespread. The exact dynamics of Cluster C are still not evident, but if this cluster represents a contrarian viewpoint, it may be more isolated or less central than the others.

### 4. Impact of Community Size Environment (Local, Internet)

The network now shows dense connections that likely cross the local and internet divides, suggesting that the messages from the superspreaders have reached widely throughout the network. There might still be differences in how local and internet nodes are integrated into the overall network, with local nodes potentially forming tighter subnetworks due to proximitybased interactions.

Looking at the InDegree vs OutDegree Heatmap at Timestep 100 The heatmap indicates a wide range of indegrees and outdegrees, with some nodes having high indegrees, suggesting that they are significant receivers of information. The network is likely to have evolved towards a few hubs with a high degree of control over the information flow.

Considering the Cumulative Distribution Function (CDF) plot The CDFs for the superspreaders show that their opinions have a significant impact on the network. The steepness of the curve at higher values indicates that the superspreaders are likely holding extreme opinions, which could polarize the network if these opinions are at odds with one another.

This analysis gives a picture of the network at a mature stage, where the effects of the superspreaders are fully integrated into the network's dynamics. The final state shows a network possibly characterized by a few strong opinion leaders and a community that may be polarized around these figures. The exact nature of consensus and the distribution of opinions would be influenced by the attributes of the nodes, their connectedness, and the strength and direction of the opinions held by the superspreaders.

## 9. Super Spreader Behavior:Opinion Dynamics Simulation

**Initialization:**

$N = 100$: Number of nodes in the network.

The network is represented as a directed graph $G$.

Each node in $G$ is assigned various attributes like opinion, stubbornness, sensitivity to external influence, trust levels towards super spreaders, community size, social influence, and distance (internet or local).

**Node Attributes:**

*Opinion*: A random value between −100 and 100.

*Stubbornness*: A random value between 0 and 1.

*External Influence Sensitivity*: A random value between 0 and 1.

*Trust in Super Spreader A/B*: A random value between 0 and 1.

*Community Size*: Randomly chosen from 'small', 'medium', or 'large'.

*Social Influence*: A random value between 0 and 1.

*Distance*: Randomly chosen from 'internet' or 'local'.

**Link Establishment:**

Links between nodes are created with a probability of 0.05, ensuring $i \neq j$.

**Super Spreader Behavior:**

Three super spreaders (A, B, C) are introduced, each with distinct opinion update logics based on Z-scores, opinion densities, and trust levels.

**Opinion Update Functions:**

Each super spreader influences neighboring nodes' opinions based on predefined Z-scores and opinion densities.

General nodes update their opinions based on the average opinion of their predecessors, modulated by their stubbornness.

**Simulation Execution:**

The opinion update process is iterated for a defined number of steps.

**Visualization:**

Node opinions are visualized using a color map ranging from blue (negative opinion) to red (positive opinion).

The network layout is generated using a spring layout algorithm.

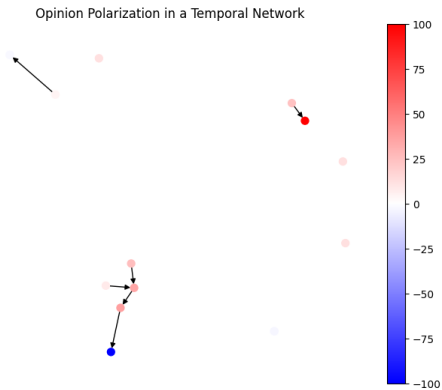

Fig. 32: Opinion Polarization in a Temporal Network $n = 10$

## 9.1 n=10 Cases

Analyzing the network at n=10 timesteps, considering the images provided, and the model description for a temporal network with opinion polarization, we can infer the following

### 1. Social Phenomenon Observation

The network diagram at n=10 illustrates a nascent stage of opinion polarization, with a few nodes starting to show distinct opinions (as indicated by the color scale). The network is not densely connected, which suggests that the spread of influence from the superspreaders is still in its initial phase.

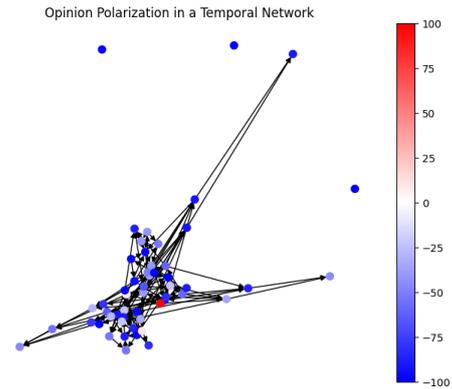

Fig. 33: Opinion Polarization in a Temporal Network $n = 50$

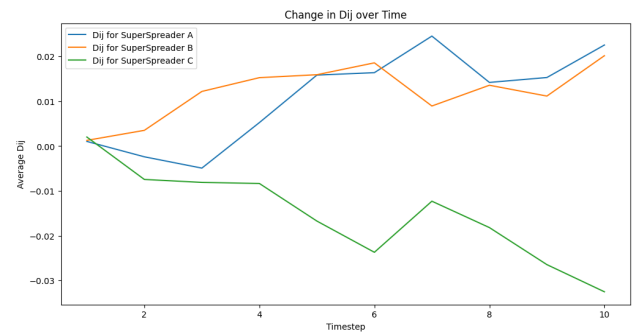

Fig. 34: Change in $Dij$ over Time, super-spreaders $A,B,C$

### 2. Consensus Formation

There is no clear consensus at this stage, as indicated by the different colors (opinions) present in the network. The lack of dense connections and the presence of isolated nodes or small subgroups point to a fragmented opinion landscape, where individuals or small groups may hold distinct views.

### 3. Cluster (A, B, C, D, E) Observation

Clusters around superspreaders A and B can be seen, with superspreader A having a direct influence on a few nodes, and superspreader B having a slightly wider reach. There is no visible cluster around superspreader C, which could suggest that its influence is not significant at this point or that it represents a contrarian opinion that is not widely shared.

### 4. Impact of Community Size Environment (Local, Internet)

At n=10, the impact of community size environment is not explicitly visible in the provided images. However, based on the model, we might expect that nodes in 'local' communities would start to show some clustering due to their propensity to be influenced by nearby nodes, whereas 'internet' nodes might be influenced by a broader array of sources.

### 5. Spreader A, B, C Dij Trust Score Evolution

The line chart shows the average Dij (difference in opinions) over time for superspreaders A, B, and C. Superspreader A maintains a relatively stable and positive Dij, which could indicate a consistent and growing trust or alignment of opinion with its followers. Superspreader B shows a fluctuating Dij, suggesting variable trust levels or alignment with its opinions over time. Superspreader C's Dij appears to decline, possibly reflecting a loss of trust or diverging opinions from the rest of the network.

The Dij scores can also be influenced by the nodes' stubbornness and external influence sensitivity, which could lead to changes in the trust or perceived reliability of the superspreaders.

Considering the model's complexity, with multiple node attributes and dynamic link establishment, the network's evolution likely involves intricate interactions between these attributes. The simulation captures the gradual process of opinion formation and the role of superspreaders in influencing this process within a temporal network.

## 9.2 n=50 Cases

## 9.3 1. Social Phenomenon Observation

The network at n=50 exhibits a more pronounced polarization of opinions. There are clearly identifiable clusters, with the one centered around a blue superspreader node suggesting a significant concentration of similar opinions. This represents a stage in the social network where influential entities have successfully swayed a large portion of the network to their opinion.

## 9.4 2. Consensus Formation

The presence of a dominant cluster indicates that a consensus may be forming around the views represented by the central superspreader. However, the existence of outlying nodes with contrasting opinions (red nodes) suggests that there is still some division within the network. The consensus is not networkwide, and there are pockets of resistance or alternative viewpoints.

## 9.5 3. Cluster (A, B, C) Observation

Without specific labels for clusters A, B, C, D, and E, it's challenging to comment on each one distinctly. However, the central cluster's density implies strong internal consensus and influence. The peripheral clusters or isolated nodes may represent minority opinions or less influential clusters.

## 9.6 4. Impact of Community Size Environment (Local, Internet)

The model does not provide explicit visualization for local versus internet nodes. Nevertheless, the impact of community size might be inferred by the density of the central cluster, which could suggest a tighter, possibly local community, versus the sparser connections that might represent broader internetbased interactions.

## 9.7 5. Spreader A, B, C, D, E Dij Trust Score Evolution

The chart tracking the change in Dij over time for superspreaders A, B, and C shows varying levels of trust or opinion alignment. Superspreader A has an increasing Dij, indicating growing trust or alignment. Superspreader B shows variability, and Superspreader C indicates a decrease in trust or diverging opinions.

The network structure and the changes in Dij scores reflect the complex interplay of individual node attributes and the evolving dynamics of influence within the network. The results show the gradual establishment of dominant opinion clusters and the shifting trust levels towards various superspreaders as the simulation progresses.

It is important to note that the analysis is based on visual interpretation of the images and the provided model description. For a more accurate assessment, a detailed examination of the network data and dynamics using computational analysis would be necessary.

## 9.8 n=100 Cases

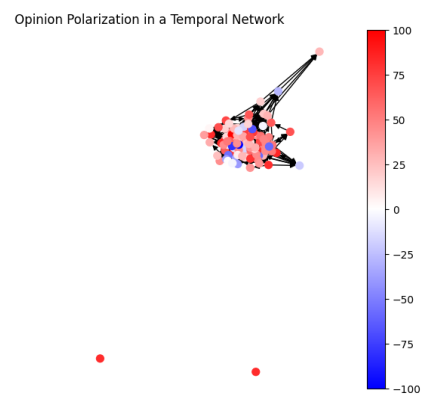

Fig. 35: Opinion Polarization in a Temporal Network $n = 100$

### 1. Social Phenomenon Consideration

With $N = 100$, the network represents a smaller community which could be akin to a niche online group or a tight-knit community in a local environment. The smaller size may lead to faster dissemination of opinions but also a quicker formation of echo chambers due to the reduced number of nodes. In the opinion polarization visualization for $N = 100$, we see that there are distinct nodes with strong opinions (red and blue) which are not closely tied to the main cluster. This indicates the existence of outliers or fringe groups that are significantly polarized compared to the central cluster.

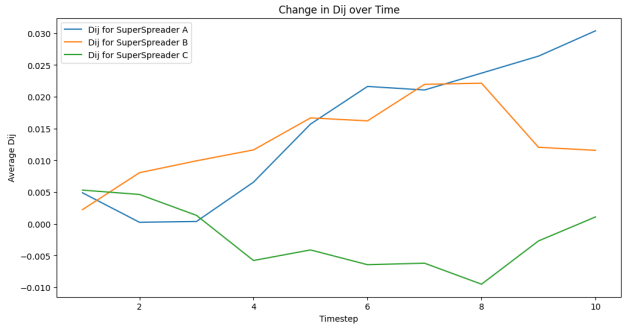

Fig. 36: Change in $Dij$ over Time, super-spreaders A,B,C

## 2. Consensus Formation

Results showing the change in $Dij$ for Super Spreaders A, B, and C suggests that the trust dynamics are more volatile in a smaller network. Super Spreader A shows a fluctuating influence, which could be a result of the smaller network size amplifying the impact of individual node attributes. Super Spreader B seems to be gaining influence more steadily, while Super Spreader C's influence varies. In such a network, consensus may be reached on a local scale, but the overall consensus may be harder to achieve due to the volatility of influence.

## 3. Cluster Consideration for A, B, C, D, E

Given the smaller size of the network, the influence of each super spreader and the dynamics within their respective clusters might be more pronounced. Each cluster's reaction to the super spreaders' opinions would likely be stronger, either leading to more cohesive clusters or more significant polarization within clusters, as minor shifts in opinion could have more noticeable effects.

## 4. Impact of Community Size and Environment (Local vs. Internet)

In a network of $N = 100$, the impact of community size and environment can be more directly observed. Smaller communities might show more rapid changes in opinion, while larger ones might resist change due to a variety of opposing views. The distinction between local and internet environments could be more pronounced as well, with local networks potentially showing stronger interpersonal influence and internet environments exhibiting more widespread but less intense interactions.

## 5. Dij Trust Score Evolution for Super Spreaders A, B, C, D, E

The evolution of $Dij$ trust scores in the provided plot indicates that the perceived reliability of super spreaders can shift more dramatically in a smaller network. Super Spreader B appears to gain trust over time, while A and C show more variability. The absence of trends for D and E in the plot suggests they either do not exist or were not included in this analysis. In a smaller network, the impact of a super spreader's trust score is likely more significant, as each node carries more weight in the network's opinion dynamics.

## 9.9 n=500 Cases

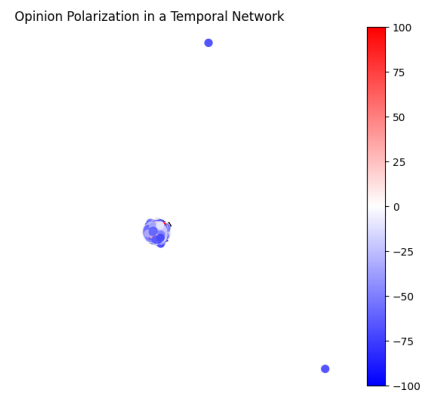

Fig. 37: Opinion Polarization in a Temporal Network $n = 500$

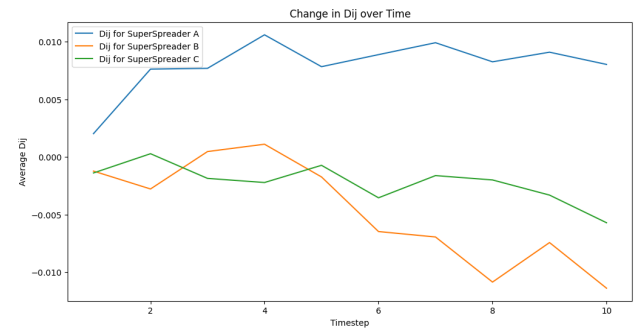

Fig. 38: Change in $Dij$ over Time, super-spreaders A,B,C

## 1. Social Phenomenon Consideration

With $N = 500$, the network is quite large, which can be representative of a medium-sized social network community, like a large online forum or a small city's social media users. The diversity in opinions and the presence of super spreaders can mimic real-world dynamics where certain individuals or entities have disproportionate influence over public discourse.

The opinion polarization visible in the first image suggests that there are clusters within the network that might be highly polarized around certain opinion values. This reflects the phenomenon of echo chambers, where individuals are mostly exposed to opinions that align with their own, reinforcing their existing beliefs.

## 2. Consensus Formation

The second image depicts the change in $Dij$ values over time for Super Spreaders A, B, and C. The trend lines indicate that the influence and trust dynamics are evolving over time. For consensus formation, it would be important to note whether the $Dij$ values are converging (indicating a movement toward consensus) or diverging (indicating increasing polarization). From the graph, it seems there is no clear trend towards consensus, especially for Super Spreader C, whose influence seems to be decreasing over time.

## 3. Cluster Consideration for A, B, C, D, E

The analysis of clusters would depend on how these clusters are formed around the super spreaders and how the opinions within these clusters interact with one another. Unfortunately, the images do not provide enough detail to analyze each cluster's behavior. However, assuming that A, B, C, D, and E are different clusters or super spreaders within the network, one could infer that clusters A and B seem to be more stable or are gaining influence, while C is losing it over time.

## 4. Impact of Community Size and Environment (Local vs. Internet)

Community size can affect the spread and adoption of opinions. Smaller communities may reach consensus more quickly but can also become echo chambers. Larger communities might have more diverse opinions but could also show significant polarization. The environment (local vs. internet) also plays a role; local networks might have stronger real-world ties and possibly more influence on an individual's opinions, whereas internet communities may contribute to the rapid spread of ideas but with potentially less trust and influence unless reinforced by super spreaders or trusted nodes.

## 5. Dij Trust Score Evolution for Super Spreaders A, B, C, D, E

The trust score's evolution is a key indicator of the super spreaders' influence over time. From the second image, Super Spreader A's trust score is relatively stable and slightly increasing, indicating a consistent or growing influence. Super Spreader B experiences some volatility but generally maintains influence. In contrast, Super Spreader C's trust score is declining, suggesting a loss of influence or trust among the nodes.

# 10. Discussion

This study delves into the complexity of opinion formation in a network environment and focuses on the magnitude of influence exerted by certain entities known as super-spreaders. classified into three types, A, B, and C, these super-spreaders play a distinct role in opinion formation throughout the network A, B, and C, play a distinct role in shaping the opinions of the network as a whole. Super-spreader A emerges as the dominant force and is characterized by a high z-score indicating that it has significant influence, especially in dense areas of opinion, such as local communities and online spaces. This entity is able to sway opinion even with potentially misleading information, thereby creating filter bubbles and echo chambers.

Conversely, super-spreader B has a low z-score and opinion density and acts as a counterbalance to A. B influences opinion in the opposite direction to A and acts as a mitigating factor for the echo chambers and filter bubbles typically fostered by A.

Super-spreader C, with a high Z-score but low opinion density, plays a unique role. It acts as an objective observer, disseminating third-party opinions and acting like a media outlet. It is hypothesized that this entity can strengthen or counteract the influence of A or B depending on the situation, acting as a coordinator or fact-checker within the network.

The study introduces a confidence coefficient Dij and a z variable to model the behavioral changes of these super-spreaders. The study shows how A's influence is more pronounced in communities where initial trust in its views is higher. In contrast, B's influence increases when C emphasizes its statements, decreasing A's trust in the cluster and causing a change in follower loyalty. This comprehensive analysis uses a directed network model to simulate the flow of information between different clusters, each playing a unique role. The study explores scenarios such as A's pervasive influence, B's receipt of information, C's wariness of A's influence, and E's opposition to A. By leveraging dynamic networks and speech analysis, this study reveals core areas of structural robustness and influence within the network over 1000 time steps.

In the vast realm of online communication, identifying unilateral preference (preference) patterns is critical to understanding and mitigating risks such as cyber-attacks and information deflection. This paper presents a comprehensive approach to analyze and visualize these patterns within social networks. Specifically, we use a directed network model to simulate the flow of information between clusters (groups) with different roles. The simulation considers the following scenarios

### 1. dominant cluster (A)

This cluster has a strong influence in the network and disseminates information over a wide area

### Passive cluster (B)

This is a large but passive cluster that is the primary recipient of information from A. It is the only one that can receive information from A. It is the only one that can receive information from B. It is the only one that can receive information from A. It is the only one that can receive information from B. It is the only one that can receive information from B. It is the only one that can receive information from B.

### Alert Cluster (C)

This cluster monitors A's activities and is alert to its influence. 4.

### Information Blocking Cluster (D)

This cluster restricts or blocks the flow of information to protect itself from A's influence. 5.

### Opposing clusters (E)

take a stand against A's information and try to limit its influence.

The key novelty of this study is the combination of dynamic network analysis and speech analysis to reveal the core (core) areas of structural robustness (durability) and influence within the network. 1000 time steps were used to simulate the network and track the evolution of message flow and the emergence of core-periphery structures. The visualization of the core provides a macroscopic view (the big picture) of the network topology (structure), vividly depicting the resilience of the network and the centrality (importance) of different clusters. Our method captures the expansion of the largest connected components (the largest group of nodes directly or indirectly connected to each other in the network) and identifies important nodes that may act as gatekeepers or influencers in the network.

The results of this study augment the detection of unidirectional communication patterns and provide deep insights into the underlying architecture (underlying structure) that facilitates such interactions. This research contributes to the field of network analysis by demonstrating novel applications in the context of social network simulation to help protect online communications.

## 11. Conclusion

Finally, we discuss the case of opinion dynamics, including future groups and environmental relationships such as location, with respect to the case of five group dynamics, taking into account ForgetfulnessFactor and CommunityPersistence.

**Network Initialization:**

Cluster Sizes: s = $\{A : 100, B : 100, C : 50, D : 60, E : 10\}$

Message Sizes: me = $\{A : 20, B : 100, C : 100, D : 50, E : 100\}$

Graph: $G$ = nx.DiGraph()

For each cluster, for each node:

Node Attributes:

cluster, opinion ~ $U(-100, 100)$,

stubbornness, sensitivity, social_influence ~ $U(0, 1)$,

community_size $\in$ {'small', 'medium', 'large'},

distance $\in$ {'internet', 'local'}

**Updating the Network at Each Timestep:**

Forgetfulness Factor: $f = 0.9$

Community Persistence: $p = 0.9$

For each node:

Old Opinion: $o_{\text{old}}$

New Opinion: $o_{\text{new}} = o_{\text{old}} \times f$

Opinion Change: $\Delta o = o_{\text{new}} - o_{\text{old}}$

Influence Factors:

External: $e = \text{sensitivity} \times \Delta o$,

Social: $s = \text{social\_influence} \times \Delta o$,

Forgetfulness: $g = f \times \Delta o$,

Stubbornness: $t = \text{stubbornness} \times \Delta o$

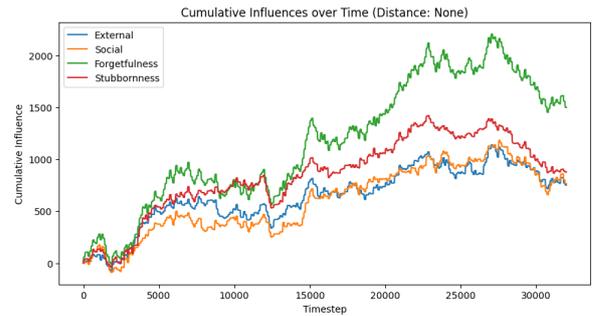

Fig. 39: Cumulative Influences over Time, Distance None

Based on the provided network initialization details and the graphs, we can discuss the behavior of the cumulative influences over time under different conditions of distance, which appears to be a factor defining the communication medium or the reachability of the nodes in the network (like 'internet', 'local', or none specified). The conditions also specify the community sizes and how they may affect forgetfulness, stubbornness, sensitivity, and social influence within the network.

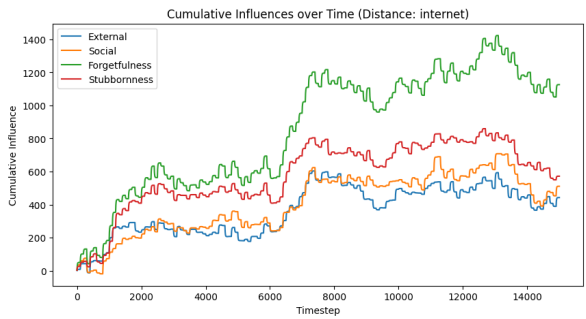

Fig. 40: Cumulative Influences over Time, Distance Internet

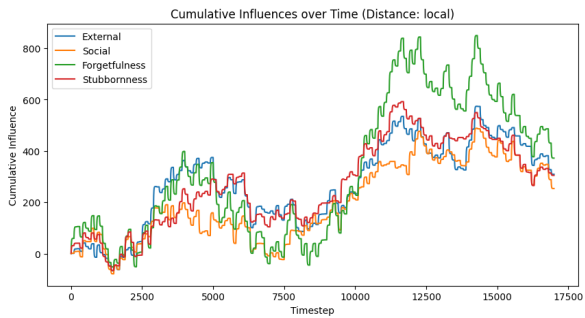

Fig. 41: Cumulative Influences over Time, Distance Local

### 1. Cumulative Influences over Time (Distance None)

This graph shows all four influences (External, Social, Forgetfulness, Stubbornness) growing over time, with Social influence being the dominant factor. This suggests that in the absence of a specified distance, social factors have the most significant impact on the network's opinions. The External influence also grows but seems to be outpaced by the Social influence, indicating that while external factors are important, the network's opinion dynamics are more heavily driven by internal social interactions. Forgetfulness and Stubbornness follow a similar trend but at a lower scale, suggesting that they are less influential than the other two factors in this setting.

### 2. Cumulative Influences over Time (Distance Local)

The presence of 'local' distance constraint seems to reduce the scale of influence when compared to the 'none' scenario. This could be due to the limited reachability or interaction between nodes in a 'local' setting. Again, Social influence is a leading factor but with less distinction from the others. This indicates that in local settings, while social influence is significant, other factors are comparably influential. The Stubbornness influence appears to be more pronounced in the 'local' setting compared to 'none', potentially indicating that in closer (local) communities, individual resistance to change plays a more significant role.

### 3. Cumulative Influences over Time (Distance Internet)

When the distance is 'internet', the scales of influence are between those of 'none' and 'local'. The Social influence is still dominant but not as distinct from External influence as in the 'none' scenario. The internet setting appears to mitigate Forgetfulness and amplify Social and External influences, possibly due to the constant influx of new information and interactions that the internet facilitates. The Stubbornness trend is similar to the 'none' scenario but with a slightly higher peak, which may suggest that while the internet allows for widespread influence, individuals still retain a degree of resistance to changing opinions.

### Community Sizes and Message Sizes

Larger communities (e.g., A and B with 100 nodes) may have a greater social influence simply due to the number of interactions possible within the community. This is reflected in the dominant Social influence in all scenarios. The message sizes, representing the number of interactions or the potential for influence, are also varying, with some clusters (e.g., B, C, and E) having a higher potential for message spread. This could result in higher cumulative social influence, as seen in the graphs.

### Network Dynamics

The initialization sets the foundation for how the network behaves. The uniform distribution of opinions, stubbornness, sensitivity, and social influence ensures a diverse set of initial conditions. The updating of the network at each timestep with a forgetfulness factor (f) and a community persistence (p) indicates that opinions are intended to decay over time unless reinforced. This decay is modulated by the forgetfulness factor and affects how the influences are calculated and how they accumulate.

In summary, the community size and the distance factor significantly affect the four attributes of forgetfulness, stubbornness, sensitivity, and social influence. The 'local' setting seems to promote stubbornness, the 'internet' enhances social and external influences, and the 'none' setting results in the highest social influence. These patterns suggest that the medium of interaction in a network can significantly shape the dynamics of opinion formation and influence spread.

### 11.1 Forgetfulness Influence

Analyzing the Forgetfulness influence heatmaps for clusters A to E, we consider the cluster sizes and the message sizes mentioned in the network conditions to understand the dynamics better. Here are the key points of analysis

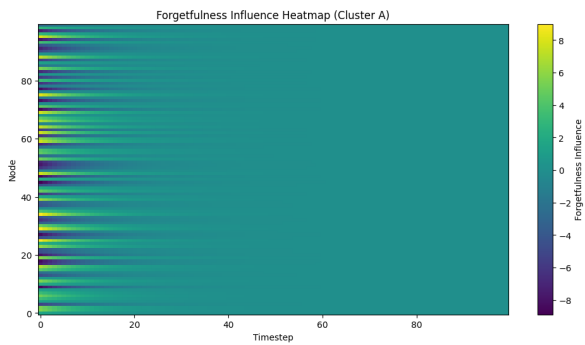

Fig. 42: Forgetfulness Influence Heatmap:A

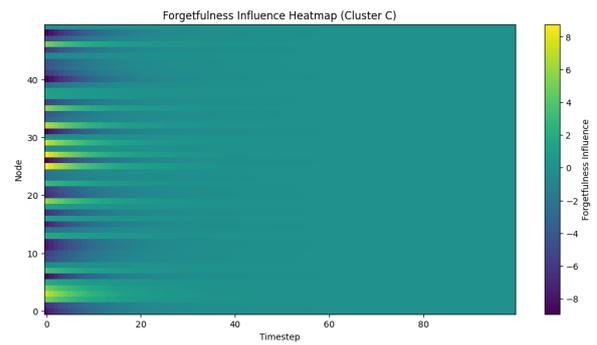

Fig. 44: Forgetfulness Influence Heatmap:C

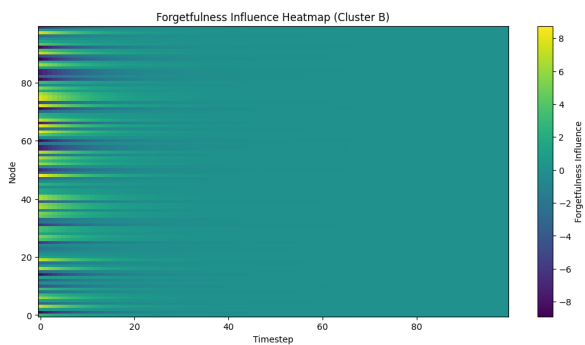

Fig. 43: Forgetfulness Influence Heatmap:B

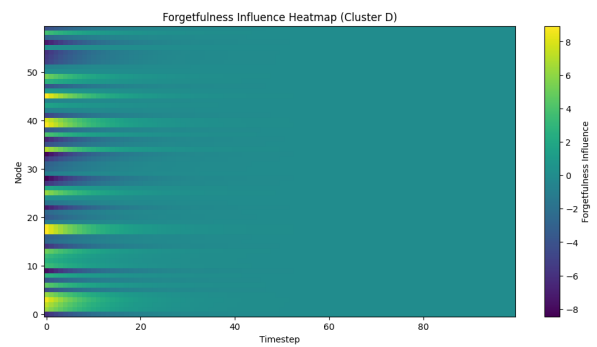

Fig. 45: Forgetfulness Influence Heatmap:D

### 1. Cluster A (100 nodes, Message size 20)

This heatmap shows a consistent pattern where a few nodes (around the 6080 range) are particularly influenced by Forgetfulness. The limited message size may contribute to this, as fewer messages could mean less reinforcement of opinions, leading to greater forgetfulness.

### 2. Cluster E (10 nodes, Message size 100)

Despite being the smallest cluster, the message size is large. This means that each node is subject to a lot of messaging, which could reinforce opinions and lead to less Forgetfulness. This is somewhat reflected in the heatmap, where we see less intense Forgetfulness influence (fewer yellow streaks) compared to larger clusters.

### 3. Cluster D (60 nodes, Message size 50)

This heatmap shows a moderate level of Forgetfulness influence. The cluster size and message size are balanced, resulting in a mix of influences. There are some nodes with higher forgetfulness, but the effect seems to be contained within specific nodes rather than widespread.

### 4. Cluster C (50 nodes, Message size 100)

With a high message size relative to the number of nodes, we might expect less forgetfulness. However, the heatmap shows several nodes with a strong Forgetfulness influence. This could indicate that other factors, such as node attributes or network structure, have a significant effect.

### 5. Cluster B (100 nodes, Message size 100)

Here we have a large cluster with a high message size, which could lead to a strong reinforcement of opinions. The heatmap shows some nodes with high Forgetfulness influence, which suggests that despite the number of messages, these nodes are influenced by the Forgetfulness factor significantly.

### 6. Overall Observations

The heatmaps show a varying degree of Forgetfulness across different clusters. The Forgetfulness influence is not uniform, indicating that nodespecific attributes and possibly the topology of the network have a significant impact. The presence of both positive and negative influences of Forgetfulness across the clusters suggests that nodes are varying in their tendency to either strengthen or weaken their opinions over time. The nodes with higher levels of Forgetfulness influence might have attributes that predispose them to change their opinion

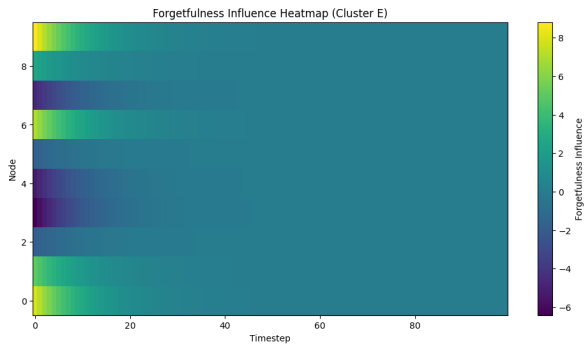

Fig. 46: Forgetfulness Influence Heatmap:E

more frequently, while nodes with lower levels might be more resilient to change.

## 7. Influence of Network Initialization Parameters

The differences across the clusters in the heatmaps could be a result of the initialization parameters such as stubbornness, sensitivity, and social influence, which are uniformly distributed. These intrinsic node properties interact with the forgetfulness factor and message dynamics to produce the observed patterns.

The Forgetfulness factor (f = 0.9) suggests that nodes retain 90% of their opinion from one timestep to the next. This retention is affected by both internal attributes and the communication dynamics within the network. The visualization shows that Forgetfulness is an active process in the network, with some nodes and timesteps showing greater influence than others.

### 11.2 Stubbornness Influence

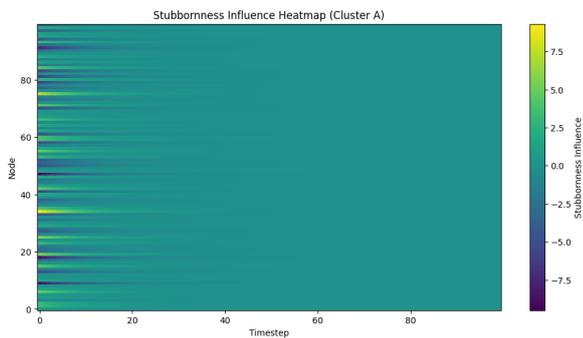

Fig. 47: Stubbornness Influence Heatmap:A

The Stubbornness influence heatmaps display how the stubbornness attribute affects each node over time within the clusters. The color scale indicates that higher values (yellow to green) suggest a stronger stubbornness influence, while lower values (purple) suggest a weaker stubbornness influence.

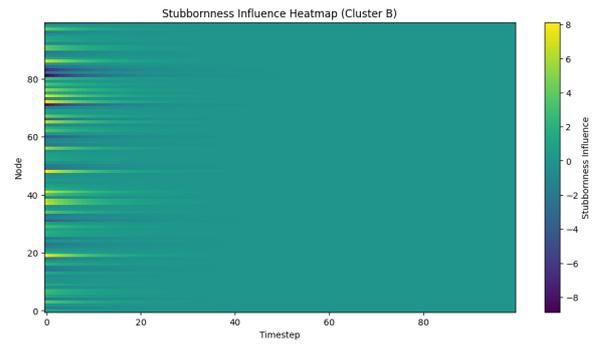

Fig. 48: Stubbornness Influence Heatmap:B

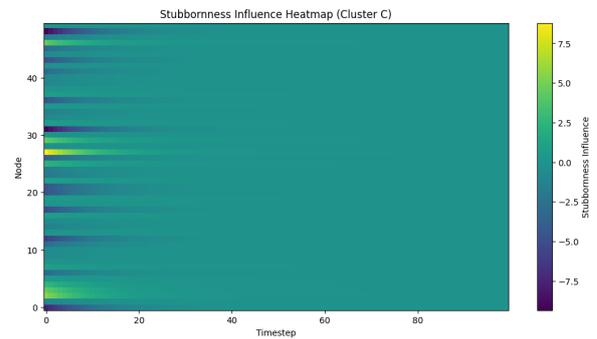

Fig. 49: Stubbornness Influence Heatmap:C

### 1. Cluster A (100 nodes, Message size 20)

This heatmap would show the influence of stubbornness in a large cluster with relatively few messages per node. The nodes with the highest levels of stubbornness influence are likely those with higher stubbornness attribute values. The limited messaging might not significantly affect these nodes' opinions due to their inherent stubbornness.

### 2. Cluster B (100 nodes, Message size 100)

With a high message size, the stubbornness influence in this cluster could be balanced by the number of messages influencing each node. Nodes with a high stubbornness attribute would maintain their opinions despite the high volume of messages, which would be indicated by more intense colors in the heatmap.

### 3. Cluster C (50 nodes, Message size 100)

This heatmap would reveal the effect of stubbornness in a mediumsized cluster with many messages per node. The stubbornness influence may be less visible due to the high message size, as frequent messaging could override some of the natural stubbornness tendencies.

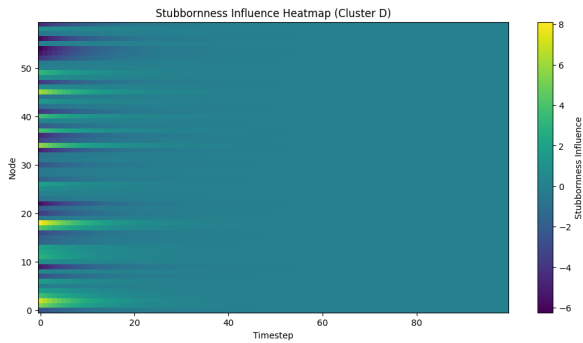

Fig. 50: Stubbornness Influence Heatmap:D

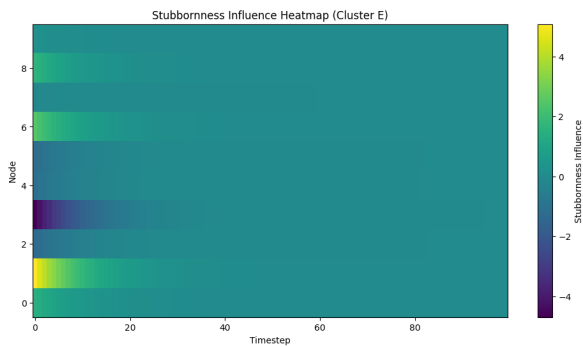

Fig. 51: Stubbornness Influence Heatmap:E

### 4. Cluster D (60 nodes, Message size 50)

The heatmap for this cluster would reflect the stubbornness influence among a moderate number of nodes with a balanced message size. We might expect a varied stubbornness influence, with some nodes showing stronger stubbornness than others.

### 5. Cluster E (10 nodes, Message size 100)

This small cluster with a high message size would typically show a minimal stubbornness influence, as the few nodes are highly interconnected with a lot of messaging. However, nodes with a high stubbornness attribute would still show a strong influence despite the messaging volume.

Across all clusters, the variability in the heatmaps reflects the diversity of stubbornness among nodes. The stubbornness attribute causes nodes to resist opinion changes regardless of the forgetfulness factor and the volume of messages they receive. This resistance to change would be reflected in the heatmaps as streaks of intense color that persist across timesteps, indicating nodes that maintain their initial opinions strongly over time.

It's important to note that the nodes' stubbornness also interacts with other factors, such as sensitivity and social influence. For instance, a node with high stubbornness but also high sensitivity might still be influenced by external changes, leading to more nuanced patterns on the heatmap.

Given the heatmaps, we would expect to see some nodes maintain consistent stubbornness levels throughout the simulation, resisting changes in their opinions. At the same time, nodes with lower stubbornness may display more significant fluctuations in opinion, appearing as changes in color intensity on the heatmap. The overall pattern would likely show clusters of nodes with similar levels of stubbornness influence, reflecting the clusters' internal structure and the nodes' connectivity.

## Aknowlegement

This research is supported by Grant-in-Aid for Scientific Research Project FY 2019-2021, Research Project/Area No. 19K04881, "Construction of a new theory of opinion dynamics that can describe the real picture of society by introducing trust and distrust". It is with great regret that we regret to inform you that the leader of this research project, Prof. Akira Ishii, passed away suddenly in the last term of the project. Prof. Ishii was about to retire from Tottori University, where he was affiliated with at the time. However, he had just presented a new basis in international social physics, complex systems science, and opinion dynamics, and his activities after his retirement were highly anticipated. It is with great regret that we inform you that we have to leave the laboratory. We would like to express our sincere gratitude to all the professors who gave me tremendous support and advice when We encountered major difficulties in the management of the laboratory at that time.

First, Prof. Isamu Okada of Soka University provided valuable comments and suggestions on the formulation of the three-party opinion model in the model of Dr. Nozomi Okano's (FY2022) doctoral dissertation. Prof.Okada also gave us specific suggestions and instructions on the mean-field approximation formula for the three-party opinion model, Prof.Okada's views on the model formula for the social connection rate in consensus building, and his analytical method. We would also like to express our sincere gratitude for your valuable comments on the simulation of time convergence and divergence in the initial conditions of the above model equation, as well as for your many words of encouragement and emotional support to our laboratory.

We would also like to thank Prof.Masaru Furukawa of Tottori University, who coordinated the late Prof.Akira Ishii's laboratory until FY2022, and gave us many valuable comments as an expert in magnetized plasma and positron research.

In particular, we would like to thank Prof.Hidehiro Matsumoto of Media Science Institute, Digital Hollywood University. Prof.Hidehiro Matsumoto is Co-author of this paper, for managing the laboratory and guiding us in the absence of the main researcher, and for his guidance on the elements of the final research that were excessive or insufficient with Prof.Masaru Furukawa.

And in particular, Prof.Masaru Furukawa of Tottori University, who is an expert in theoretical and simulation research on physics and mathematics of continuum with a focus on magnetized plasma, gave us valuable opinions from a new perspective.

His research topics include irregular and perturbed magnetic fields, MHD wave motion and stability in non-uniform plasmas including shear flow, the boundary layer problem in magnetized plasmas, and pseudo-annealing of MHD equilibria with magnetic islands.

We received many comments on our research from new perspectives and suggestions for future research. We believe that Prof.Furukawa's guidance provided us with future challenges and perspectives for this research, which stopped halfway through. We would like to express sincere gratitude to him.

We would like to express my sincere gratitude to M Data Corporation, Prof.Koki Uchiyama of Hotlink Corporation, Prof.Narihiko Yoshida, President of Hit Contents Research Institute, Professor of Digital Hollywood University Graduate School, Hidehiko Oguchi of Perspective Media, Inc. for his valuable views from a political science perspective. And Kosuke Kurokawa of M Data Corporation for his support and comments on our research environment over a long period of time. We would like to express our gratitude to Hidehiko Oguchi of Perspective Media, Inc. for his valuable views from the perspective of political science, as well as for his hints and suggestions on how to build opinion dynamics.

We are also grateful to Prof.Masaru Nishikawa of Tsuda University for his expert opinion on the definition of conditions in international electoral simulations.

We would also like to thank all the Professors of the Faculty of Engineering, Tottori University. And Prof.Takayuki Mizuno of the National Institute of Informatics, Prof.Fujio Toriumi of the University of Tokyo, Prof.Kazutoshi Sasahara of the Tokyo Institute of Technology, Prof.Makoto Mizuno of Meiji University, Prof.Kaoru Endo of Gakushuin University, and Prof.Yuki Yasuda of Kansai University for taking over and supporting the Society for Computational Social Sciences, which the late Prof.Akira Ishii organized, and for their many concerns for the laboratory's operation. We would also like to thank Prof.Takuju Zen of Kochi University of Technology and Prof.Serge Galam of the Institut d'Etudes Politiques de Paris for inviting me to write this paper and the professors provided many suggestions regarding this long-term our research projects.

We also hope to contribute to their further activities and the development of this field. In addition, we would like to express our sincere gratitude to Prof.Sasaki Research Teams for his heartfelt understanding, support, and advice on the content of our research, and for continuing our discussions at a time when the very survival of the research project itself is in jeopardy due to the sudden death of the project leader.

We would also like to express our sincere gratitude to the bereaved Family of Prof.Akira Ishii, who passed away unexpectedly, for their support and comments leading up to the writing of this report. We would like to close this paper with my best wishes for the repose of the soul of Prof.Akira Ishii, the contribution of his research results to society, the development of ongoing basic research and the connection of research results, and the understanding of this research project.

# References

zh

[Dal Pozzolo, A., Boracchi, G., Caelen, O(2018)] Dal Pozzolo, A., Boracchi, G., Caelen, O., Alippi, C., Bontempi, G.. Credit Card Fraud Detection: A Realistic Modeling and a Novel Learning Strategy. *IEEE transactions on neural networks and learning systems*, **2018**.

[Buczak, A. L., Guven, E.(2016)] Buczak, A. L., Guven, E.. A Survey of Data Mining and Machine Learning Methods for Cyber Security Intrusion Detection. *IEEE Communications Surveys Tutorials*, **2016**.

[Alpcan, T., Başar, T.(2006)] Alpcan, T., Başar, T.. An Intrusion Detection Game with Limited Observations. *12th International Symposium on Dynamic Games and Applications*,**2006**.

[Schlegl, T., Seebock, P., Waldstein, S. M., Schmidt-Erfurth, U., Langs, Schlegl, T., Seebock, P., Waldstein, S. M., Schmidt-Erfurth, U., Langs, G.. Unsupervised Anomaly Detection with Generative Adversarial Networks to Guide Marker Discovery. *Information Processing in Medical Imaging*, **2017**.

[Mirsky, Y., Doitshman, T., Elovici, Y., Shabtai, A.(2018)] Mirsky, Y., Doitshman, T., Elovici, Y., Shabtai, A.. Kitsune: An Ensemble of Autoencoders for Online Network Intrusion Detection. *Network and Distributed System Security Symposium*,**2018**.

[Alpcan, T., Başar, T.(2003)] Alpcan, T., Başar, T.. A Game Theoretic Approach to Decision and Analysis in Network Intrusion Detection. *Proceedings of the 42nd IEEE Conference on Decision and Control* ,**2003**.

[Nguyen, K. C., Alpcan, T., Başar, T.(2009)] Nguyen, K. C., Alpcan, T., Başar, T.. Stochastic Games for Security in Networks with Interdependent Nodes. *International Conference on Game Theory for Networks*,**2009**.

[Tambe, M.(2011)] Tambe, M.. Security and Game Theory: Algorithms, Deployed Systems, Lessons Learned. *Cambridge University Press* **[Volume]**,**2011**.

[Korilis, Y. A., Lazar, A. A., Orda, A.(1997)] Korilis, Y. A., Lazar, A. A., Orda, A.. Achieving Network Optima Using Stackelberg Routing Strategies. *IEEE/ACM Transactions on Networking*,**2011**.

[Hausken, K.(2013)] Hausken, K.. Game Theory and Cyber Warfare. *The Economics of Information Security and Privacy*,**2013**.

[Justin, S., et al.(2020)] Justin, S., et al.. Deep learning for cyber security intrusion detection: Approaches,

[datasets, and comparative study. *Journal of Information Security and Applications, vol. 50*, **2020**.

[Zenati, H., et al.(2018)] Zenati, H., et al.. Efficient GAN-Based Anomaly Detection. *Workshop Track of ICLR*,**2018**.

[Roy, S., et al.(2010)] Roy, S., et al.. A survey of game theory as applied to network security. *43rd Hawaii International Conference on System Sciences*,**2010**.

[Biggio, B., Roli, F.(2018)] Biggio, B., Roli, F.. Wild patterns: Ten years after the rise of adversarial machine learning. *Pattern Recognition, vol. 84*,**2018**.

[Massanari2017] Massanari, A. (2017). #Gamergate and The Fappening: How Reddit's algorithm, governance, and culture support toxic technocultures. *New Media & Society*, **2017**, *19*(3), 329-346.

[Castells2012] Castells, M. (2012). Networks of Outrage and Hope: Social Movements in the Internet Age. Polity Press, **2012**.

[Wojcieszak(2010)] Wojcieszak, M. 'Don't talk to me': Effects of ideologically homogeneous online groups and politically dissimilar offline ties on extremism. *New Media & Society* **2010**, *12*(4), pp.637-655.

[Tucker et al.(2017)] Tucker, J. A.; Theocharis, Y.; Roberts, M. E.; Barberá, P. From Liberation to Turmoil: Social Media And Democracy. *Journal of Democracy* **2017**, *28*(4), pp.46-59.

[Conover et al.(2011)] Conover, M. D.; Ratkiewicz, J.; Francisco, M.; Gonçalves, B.; Menczer, F.; Flammini, A. Political polarization on Twitter. In Proceedings of the ICWSM, **2011**; Vol. 133, pp.89-96.

[Chen & Wellman(2004)] Chen, W.; Wellman, B. The global digital divide – within and between countries. *IT & Society* **2004**, *1*(7), pp.39-45.

[Van Dijck(2013)] Van Dijck, J. *The Culture of Connectivity: A Critical History of Social Media*; Oxford University Press: Oxford, UK, **2013**.

[Bakshy et al.(2015)] Bakshy, E.; Messing, S.; Adamic, L. A. Exposure to ideologically diverse news and opinion on Facebook. *Science* **2015**, *348*(6239), pp.1130-1132.

[Jost et al.(2009)] Jost, J. T.; Federico, C. M.; Napier, J. L. Political ideology: Its structure, functions, and elective affinities. *Annual Review of Psychology* **2009**, *60*, pp.307-337.

[Iyengar & Westwood(2015)] Iyengar, S.; Westwood, S. J. Fear and loathing across party lines: New evidence on group polarization. *American Journal of Political Science* **2015**, *59*(3), pp.690-707.

[Green et al.(2002)] Green, D. P.; Palmquist, B.; Schickler, E. *Partisan Hearts and Minds: Political Parties and the Social Identities of Voters*; Yale University Press: New Haven, CT, **2002**.

[McCoy et al.(2018)] McCoy, J.; Rahman, T.; Somer, M. Polarization and the Global Crisis of Democracy: Common Patterns, Dynamics, and Pernicious Consequences for Democratic Polities. *American Behavioral Scientist* **2018**, *62*(1), pp.16-42.

[Tucker et al.(2018)] Tucker, J. A., et al. Social Media, Political Polarization, and Political Disinformation: A Review of the Scientific Literature. SSRN, **2018**.

[Bail(2020)] Bail, C. A. *Breaking the Social Media Prism: How to Make Our Platforms Less Polarizing*; Princeton University Press: Princeton, NJ, **2020**.

[Barberá(2015)] Barberá, P. Birds of the Same Feather Tweet Together: Bayesian Ideal Point Estimation Using Twitter Data. *Political Analysis* **2015**, *23*(1), pp.76-91.

[Garimella et al.(2018)] Garimella, K., et al. Political Discourse on Social Media: Echo Chambers, Gatekeepers, and the Price of Bipartisanship. In Proceedings of the 2018 World Wide Web Conference on World Wide Web, **2018**.

[Allcott & Gentzkow(2017)] Allcott, H.; Gentzkow, M. Social Media and Fake News in the 2016 Election. *Journal of Economic Perspectives* **2017**, *31*(2), pp.211-236.

[Garrett(2009)] Garrett, R. K. Echo Chambers Online?: Politically Motivated Selective Exposure among Internet News Users. *Journal of Computer-Mediated Communication* **2009**, *14*(2), pp.265-285.

[Weeks & Cassell(2016)] Weeks, B. E.; Cassell, A. Partisan Provocation: The Role of Partisan News Use and Emotional Responses in Political Information Sharing in Social Media. *Human Communication Research* **2016**, *42*(4), pp.641-661.

[Iyengar et al.(2012)] Iyengar, S.; Sood, G.; Lelkes, Y. Affect, Not Ideology: A Social Identity Perspective on Polarization. *Public Opinion Quarterly* **2012**, *76*(3), pp.405-431.

[Bimber(2014)] Bimber, B. Digital Media in the Obama Campaigns of 2008 and 2012: Adaptation to the Personalized Political Communication Environment. *Journal of Information Technology & Politics* **2014**,

[Castellano et al.(2009)] Castellano, C., Fortunato, S., & Loreto, V. Statistical physics of social dynamics. *Reviews of Modern Physics* **2009**, *81*, 591-646.

[Sîrbu et al.(2017)] Sîrbu, A., Loreto, V., Servedio, V.D.P., & Tria, F. Opinion Dynamics: Models, Extensions and External Effects. In: Loreto V. et al. (eds) Participatory Sensing, Opinions and Collective Awareness. *Understanding Complex Systems*. Springer, Cham, **2017**.

[Deffuant et al.(2000)] Deffuant, G., Neau, D., Amblard, F., & Weisbuch, G. Mixing Beliefs among Interacting Agents. *Advances in Complex Systems* **2000**, *3*, 87-98.

[Weisbuch et al.(2002)] Weisbuch, G., Deffuant, G., Amblard, F., & Nadal, J. P. Meet, Discuss and Segregate!. *Complexity* **2002**, *7*(3), 55-63.

[Hegselmann & Krause(2002)] Hegselmann, R., & Krause, U. Opinion Dynamics and Bounded Confidence Models, Analysis, and Simulation. *Journal of Artificial Society and Social Simulation* **2002**, *5*, 1-33.

[Ishii & Kawahata(2018)] Ishii A. & Kawahata, Y. Opinion Dynamics Theory for Analysis of Consensus Formation and Division of Opinion on the Internet. In: Proceedings of The 22nd Asia Pacific Symposium on Intelligent and Evolutionary Systems, 71-76, **2018**. arXiv:1812.11845 [physics.soc-ph]

[Ishii(2019)] Ishii A. Opinion Dynamics Theory Considering Trust and Suspicion in Human Relations. In: Morais D., Carreras A., de Almeida A., Vetschera R. (eds) Group Decision and Negotiation: Behavior, Models, and Support. GDN 2019. *Lecture Notes in Business Information Processing* 351, Springer, Cham 193-204, **2019**.

[Ishii & Kawahata(2019)] Ishii A. & Kawahata, Y. Opinion dynamics theory considering interpersonal relationship of trust and distrust and media effects. In: The 33rd Annual Conference of the Japanese Society for Artificial Intelligence 33. JSAI2019 2F3-OS-5a-05, **2019**.

[Agarwal et al.(2011)] Agarwal, A, Xie, B., Vovsha, I., Rambow, O. & Passonneau, R. Sentiment analysis of twitter data. In: Proceedings of the workshop on languages in social media. Association for Computational Linguistics 30-38, **2011**.

[Siersdorfer et al.(2010)] Siersdorfer, S., Chelaru, S. & Nejdl, W. How useful are your comments?: analyzing and predicting youtube comments and comment ratings. In: Proceedings of the 19th international conference on World wide web. 891-900, **2010**.

[Wilson et al.(2005)] Wilson, T., Wiebe, J., & Hoffmann, P. Recognizing contextual polarity in phrase-level sentiment analysis. In: Proceedings of the conference on human language technology and empirical methods in natural language processing 347-354, **2005**.

[Sasahara et al.(2020)] Sasahara, H., Chen, W., Peng, H., Ciampaglia, G. L., Flammini, A. & Menczer, F. On the Inevitability of Online Echo Chambers. arXiv: 1905.03919v2, **2020**.

[Ishii and Kawahata(2018)] Ishii, A.; Kawahata, Y. Opinion Dynamics Theory for Analysis of Consensus Formation and Division of Opinion on the Internet. In *Proceedings of The 22nd Asia Pacific Symposium on Intelligent and Evolutionary Systems (IES2018)*, 71-76; arXiv:1812.11845 [physics.soc-ph], **2018**.

[Ishii(2019)] Ishii, A. Opinion Dynamics Theory Considering Trust and Suspicion in Human Relations. In *Group Decision and Negotiation: Behavior, Models, and Support. GDN 2019. Lecture Notes in Business Information Processing*, Morais, D.; Carreras, A.; de Almeida, A.; Vetschera, R. (eds.), **2019**, *351*, 193-204.

[Ishii and Kawahata(2019)] Ishii, A.; Kawahata, Y. Opinion dynamics theory considering interpersonal relationship of trust and distrust and media effects. In *The 33rd Annual Conference of the Japanese Society for Artificial Intelligence*, JSAI2019 2F3-OS-5a-05, **2019**.

[Okano and Ishii(2019)] Okano, N.; Ishii, A. Isolated, untrusted people in society and charismatic person using opinion dynamics. In *Proceedings of ABCSS2019 in Web Intelligence 2019*, 1-6, **2019**.

[Ishii and Kawahata(2019)] Ishii, A.; Kawahata, Y. New Opinion dynamics theory considering interpersonal relationship of both trust and distrust. In *Proceedings of ABCSS2019 in Web Intelligence 2019*, 43-50, **2019**.

[Okano and Ishii(2019)] Okano, N.; Ishii, A. Sociophysics approach of simulation of charismatic person and distrusted people in society using opinion dynamics. In *Proceedings of the 23rd Asia-Pacific Symposium on Intelligent and Evolutionary Systems*, 238-252, **2019**.

[Ishii and Okano(2021)] Ishii, A, and Nozomi,O. Sociophysics approach of simulation of mass media effects in society using new opinion dynamics. In *Intelligent Systems and Applications: Proceedings of the 2020 Intelligent Systems Conference (IntelliSys) Volume 3*. Springer International Publishing, **2021**.

[Ishii and Kawahata(2020)] Ishii, A.; Kawahata, Y. Theory of opinion distribution in human relations where trust and distrust mixed. In Czarnowski, I., et al. (eds.), *Intelligent Decision Technologies*, Smart Innovation, Systems and Technologies 193, **2020**; pp. 471-478.

[Ishii et al.(2021)] Ishii, A.; Okano, N.; Nishikawa, M. Social Simulation of Intergroup Conflicts Using a New Model of Opinion Dynamics. *Front. Phys.* **2021**, *9*:640925. doi: 10.3389/fphy.2021.640925.

[Ishii et al.(2020)] Ishii, A.; Yomura, I.; Okano, N. Opinion Dynamics Including both Trust and Distrust in Human Relation for Various Network Structure. In *The Proceeding of TAAI 2020*, in press, **2020**.

[Fujii and Ishii(2020)] Fujii, M.; Ishii, A. The simulation of diffusion of innovations using new opinion dynamics. In *The 2020 IEEE/WIC/ACM International Joint Conference on Web Intelligence and Intelligent Agent Technology*, in press, **2020**.

[Ishii & Okano(2021)] Ishii, A, Okano, N. Social Simulation of a Divided Society Using Opinion Dynamics. *Proceedings of the 2020 IEEE/WIC/ACM International Joint Conference on Web Intelligence and Intelligent Agent Technology* (in press) **2021**.

[Ishii & Okano(2021)] Ishii, A., & Okano, N. Sociophysics Approach of Simulation of Mass Media Effects in Society Using New Opinion Dynamics. In *Intelligent Systems and Applications (Proceedings of the 2020 Intelligent Systems Conference (IntelliSys) Volume 3)*, pp. 13-28. Springer, **2021**.

[Okano & Ishii(2021)] Okano, N. & Ishii, A. Opinion dynamics on a dual network of neighbor relations and society as a whole using the Trust-Distrust model. In *Springer Nature - Book Series: Transactions on Computational Science & Computational Intelligence (The 23rd International Conference on Artificial Intelligence (ICAI'21))*, **2021**.